\documentclass[fleqn,usenatbib,twocolumn]{mnras}
\usepackage{graphicx} 
\usepackage{xcolor} 
\usepackage{hyperref}
\makeatletter\let\expandableinput\@@input\makeatother
\usepackage{newtxtext,newtxmath}
\usepackage[normalem]{ulem}

\usepackage[T1]{fontenc}
\usepackage{ae,aecompl}
\usepackage{float,comment}
\usepackage{longtable,nicefrac}
\usepackage{array}
\newcolumntype{H}{>{\setbox0=\hbox\bgroup}c<{\egroup}@{}}

\definecolor{DG}{rgb}{0.09, 0.45, 0.27}
\def\ion{\mathrm{ion}}
\def\kms{km~s$^{-1}$}

\def\Ob{$\Omega_b$}
\def\Om{$\Omega_m$}
\def\ObX{$\Omega_{\mathrm{WHIM,X}}$}
\def\ObXEq{\Omega_{\mathrm{WHIM,X}}}
\def\ObXb{$\Omega_{\mathrm{WHIM,X}}/\Omega_b$}

\def\ObXbFracEq{\dfrac{\Omega_{\mathrm{WHIM,X}}}{\Omega_b}}
\def\ObXbEq{\Omega_{\mathrm{WHIM,X}}/\Omega_b}
\def\fion{f_{\ion}}

\def\NULeq{N^{\mathrm{U}}_{\mathrm{ion}}}
\def\NUL{$\NULeq$}

\def\cog{curve--of--growth}

\def\dof{$\mathrm{d.o.f.}$}
\def\E{\text{E}}

\def\oi{O~I}
\def\oii{O~II}

\def\ovi{O~VI}
\def\ovii{O~VII}
\def\oviii{O~VIII}

\def\nv{N~V}

\def\hi{H~I}

\def\nion{N_{\mathrm{ion}}}
\def\nioni{N_{\mathrm{ion},i}}
\def\nionj{N_{\mathrm{ion,j}}}
\def\E{\mathrm{E}}
\def\chandra{\it Chandra\rm}
\def\xmm{\it XMM-Newton\rm}
\def\xmmshort{\it XMM\rm}

\def\es{1ES~1553+113}
\def\kms{km~s$^{-1}$}

\def\OWHIMX{\Omega_{\mathrm{WHIM,X}}}

\def\eagle{\texttt{EAGLE}}
\def\spex{\texttt{SPEX}}
\def\cstat{\emph{C}--stat}
\def\gof{goodness--of--fit}
\def\Wl{W_{\lambda}}
\def\spex{\texttt{SPEX}}
\def\cmin{$C_\mathrm{min}$}

\def\ml{maximum--likelihood}
\def\paperOne{paper~I}

\def\nSources{51}
\def\nsystems{1,224}
\def\nDet{33}

\usepackage{graphicx}   
\usepackage{amsmath}    
\usepackage{scalerel}
\title[Cosmology from quasar X--ray absorption]{X--ray absorption lines in FUV--detected quasars:\\
II. Cosmological density and properties  of
the \emph{missing baryons}}

\author[]{
Massimiliano Bonamente$^{1}$\thanks{E-mail: bonamem@uah.edu}, David~Spence$^{1}$, Jussi~Ahoranta$^{2}$, 
Nastasha~Wijers$^{3}$, Toni~Tuominen$^{2}$
\newauthor{and Jelle~De~Plaa$^{4}$}\\
$^{1}$ Department of Physics and Astronomy, University of Alabama in Huntsville, Huntsville, AL \\
$^{2}$ Department of Physics, University of Helsinki, PO Box 64, 00014 Helsinki, Finland \\
$^{3}$ Center for Interdisciplinary Exploration and Research in Astrophysics (CIERA) and Department of Physics and Astronomy,\\ Northwestern University, 1800 Sherman Ave, Evanston, IL 60201, USA\\
$^{4}$ SRON, Netherlands Institute for Space Research Astrophysics, Niels Bohrweg 4, 2333CA Leiden, Netherlands\\
}
\date{}

\begin{document}

\maketitle
\begin{abstract}
    This paper presents constraints on the cosmological density of baryons from a systematic
    search for \ovii\ and \oviii\  absorption lines in the \xmm\ and \chandra\ X--ray spectra of \nSources\ background sources.
    The search is based on far ultra--violet redshift priors from HST and FUSE, and it has 
    resulted in the identification of 34 possible 
    \ovii\ and \oviii\ absorption--line systems
    at the 99\% confidence level, out of a search in \nsystems\ systems with fixed redshift priors. Of these,  7 \ovii\ and 8 \oviii\ systems pass additional screening criteria
    and are deemed to be associated with
    the warm--hot intergalactic medium (WHIM). 
    We find that the cosmological baryon density associated with
    these possible detections is consistent with the value required to solve the \emph{missing baryons} problem. Specifically, we find that $\Omega_{\mathrm{WHIM,X}}/\Omega_b=0.83^{+3.99}_{-0.62}$ from the \ovii\ lines, at the 68\% level of confidence (assuming 20\% Solar abundances and 100\% ionization fraction), 
     or separately $\Omega_{\mathrm{WHIM,X}}/\Omega_b=0.79^{+3.08}_{-0.50}$ from the \oviii\ lines (assuming 20\% Solar abundances 50\% ionization fraction). 
    We also conducted an extensive analysis of systematic errors affecting these estimates, 
    and provided
    evidence of the association between the detected X--ray absorption line systems with known
    filaments of SDSS galaxies.  The results of this analysis
    therefore contributes to the characterization of the {missing baryons} and it indicates that 
    they are in fact associated with the high--temperature portion
    of the warm-hot intergalactic medium, and possibly with large-scale WHIM filaments traced by galaxies, as consistently predicted by numerical simulations and by other independent probes.
\end{abstract}


\section{Introduction}
\label{sec:introduction}

Numerical simulations consistently indicate that
 filamentary structures of galaxies are a significant reservoir of warm--hot
intergalactic medium at temperatures of approximately $\log T(K)=5-7$ \citep[the WHIM, e.g.][]{cen1999,dave2001,bertone2008,cautun2014}.
Given that a significant fraction of the present--day baryons
{  is} 
unaccounted for  {  in FUV searches} \citep[i.e., the \emph{missing baryons} problem, e.g.][]{danforth2016}, it is natural to expect that { X--ray observations of} the WHIM hold the
key to {  addressing} 
this problem.
In fact, there is growing evidence that the missing baryons are in the high--temperature
($\log T(K)\geq 6$) portion of the WHIM, and therefore primarily accessible
only via X--ray observations of such ions as \ovii\ and \oviii, and others \citep[see, e.g.][]{martizzi2019,tuominen2021}.

The higher range of WHIM temperatures has been more challenging to 
probe than  lower temperatures \citep[see, e.g.][for a recent analysis]{gatuzz2023}, primarily because the available spectrometers on board
\xmm\ and \chandra\ are far less  sensitive than the FUV instruments \citep[e.g.][]{danforth2016, tilton2012}.
The challenges associated with the low count rates and the limited resolution
in X--ray observations of background quasars has resulted in only a handful
of possible X--ray absorption line detections, primarily from \oviii\ and \ovii\
ions \citep[e.g.][]{spence2023,ahoranta2021,ahoranta2020,fang2002,bonamente2016,nicastro2018, kovacs2019}. The 
{  characterization of the} missing baryons problem therefore remains one of 
observational cosmology's most outstanding unsolved problems.

In an effort to solve this problem, \paperOne\ {  \citep{spence2024}} presented the most comprehensive analysis to date 
of all X--ray sightlines towards bright quasars observed by \xmm\ and \chandra, in search for \ovii\ and \oviii\ absorption at the redshift of prior FUV absorption lines
in the \cite{danforth2016} and \cite{tilton2012} data. In that paper,
we have identified a number of possible X--ray absorption line systems
that coincide with FUV absorption, some at the same wavelength or adjacent to previously
reported detections.
In \paperOne\ we have also set upper limits to the non--detection of absorption lines for all sources and all FUV redshift priors.
In this paper we
use the results of \paperOne\ to set upper limits to the non--detection of
the column density of \ovii\ and \oviii, to provide estimates for the 
column density of the possible detections, and to estimate the cosmological
density of baryons associated with these measurements. {  The goal of the analysis is to characterize the X--ray absorbing WHIM and its relationship to the missing baryons.}

This paper is structured as follows. 
In Sec.~\ref{sec:tau0} we briefly review the result of \paperOne, and present
upper limits to the optical depth and the column density of the lines for all systems
studied in \paperOne. In Sec.~\ref{sec:cosmo} we present the constraints
on the cosmological baryon density of the WHIM based on both detections and upper limits.
In Sec.~\ref{sec:interpretation} we discuss the cosmological interpretation of the results 
within the context of the missing baryons problem, and  Sec.~\ref{sec:discussion} contains our
conclusions.

\section{Constraints on the optical depth and column density of the absorption lines}
\label{sec:tau0}

\subsection{The X--ray sample of systems with FUV priors}
\def\line{\texttt{line}}
The sample presented in \paperOne\ and used for this study consists of \nSources\ X--ray sources observed by either the \xmm\ RGS or 
the \chandra\ LETG grating spectrometers,
with many sources observed by both, for a total of 16.2~Ms of RGS exposure and 4.1~Ms of LETG exposure.
For these sources, \cite{danforth2016} (hereafter D16) and \cite{tilton2012} (hereafter T12) 
detected 332 \ovi\ and 178 \hi\ BLA systems. The FUV redshifts were searched for \ovii\ K-$\alpha$
and \oviii\ Lyman-$\alpha$ lines, which are located at respectively 21.602~\AA\ and 18.969~\AA\ in the rest frame, by fitting a portion of the spectrum with a power--law model modified by a Gaussian \line\ model.

The \line\ model has one free parameter, the optical depth at line center $\tau_0$, which is therefore
the main observable of this study. An absorption line feature has a positive optical depth, while
an emission line feature has a negative optical depth, and the relationship
between $\tau_0$ and the column density (i.e., the \cog\ of the line) is discussed in Sec.~\ref{sec:colDen} {  and in App.~\ref{app:tables}}. The vast majority of spectral fits had an acceptable
\gof\ statistic \cmin, with just a few fits featuring a poorer fit. Certain systems could not be 
searched for X--ray lines due to a variety of reasons, such as detector artifacts or overlapping Galactic lines, as discussed in \paperOne.

Of the \nsystems\ systems studied as part of this project { (see Table~4 of \paperOne\ for criteria of selection)}, there are \nDet\ systems where absorption 
was detected at 99\% confidence level, as measured by the $\Delta C \geq 6.6$ statistic \citep[see, e.g.][]{cash1979}, plus one
additional system with significant absorption but with a
poor{er} fit. Those systems were
studied in detail in Sec.~3.4 and Table~13 of \paperOne, and include certain absorption line detections that 
are either associated with the Galaxy, or with absorbers that are likely intrinsic to the source. 
The possible WHIM  origin of these absorption--line detections is investigated in more detail in Sec~\ref{sec:interpretation} { of this paper}.

\subsection{Constraints { on} the optical depth parameter}
\label{sec:ULMethods}

For the \nDet+1 systems where absorption 
was detected at 99\% confidence level, the optical depth parameters  $\tau_0$ are already available for
analysis from Tables~5 through ~12 of \paperOne.
Given that the majority of X--ray absorption--line systems resulted in non--detections, it is
also necessary to establish upper limits for
the $\tau_0$ parameter for the systems without a detection. Such upper limits will then be used to determine upper limits
to the associated column densities, and therefore of the cosmological density of baryons.

By design, the $\tau_0$
parameter is allowed to have either sign, although
the physically allowed range for an absorption line is $\tau_0 \geq 0$, with
$\tau_0=0$ representing the null hypothesis that there is no absorption line. The reason for allowing
the parameter to be any real number is so that the $\Delta C$ \gof\ statistic for the 
presence of an absorption can be used for hypothesis testing (see discussion in Sec.~3.1 of \paperOne).
In fact, the  $\Delta C$ statistic can be used only if the null--hypothesis value (in this case $\tau_0=0$)
does \emph{not} lie at the boundaries of the allowed parameter space \citep[e.g., as was discussed in
an astrophysical context by][]{protassov2002}. 

The statistical issues associated with the determination of an upper limit, and the choices
made for this application, are discussed in Appendix~\ref{app:UL}.
We chose to use two independent methods to determine upper limits. 
The first makes use of the popular \cite{feldman1997} likelihood-ratio ordering criterion,
which naturally overcomes  the fact that many of the measurements yield
a negative best--fit $\tau_0$. This method will be referred to as the Feldman--Cousins (FC)
method, and it makes use of the assumption that absorption lines require $\tau_0\geq 0$. 
The second is a simpler method that assumes a true value of $\tau_0=0$, i.e., the absence of either emission of absorption, regardless
of the measured best--fit. It consists of simply setting the average of the positive and negative
uncertainties in the parameter  as the upper--limit to the value of the parameter (optionally multiplied
by a constant, according to the chosen level of significance of the upper limit).
For example, the best--fit value for the first entry in Table~5 of \paperOne\
is $\tau_0=+0.55\pm^{1.78}_{0.88}$, resulting in an upper limit of $1.31$. The second method will
be referred to as the `sensitivity' method.

%

\subsection{The line profile and { curves of growth}}
\label{sec:lineProfile}

The key physical parameter associated with each absorption line is the column density { $N$}
of the intervening ions. It is therefore critical to link
the model parameter $\tau_0$ to the ion column density { and to the equivalent width $\Wl$}, and to understand the main sources
of systematic uncertainties associated with this step of the analysis. 
{ The methods used in this paper follow the standard curve--of--growth analysis of \cite{draine2011}, and the  relevant equations are reported in App.~\ref{sec:cog}.}

\begin{figure*}
\centering
    \includegraphics[width=3.4in]{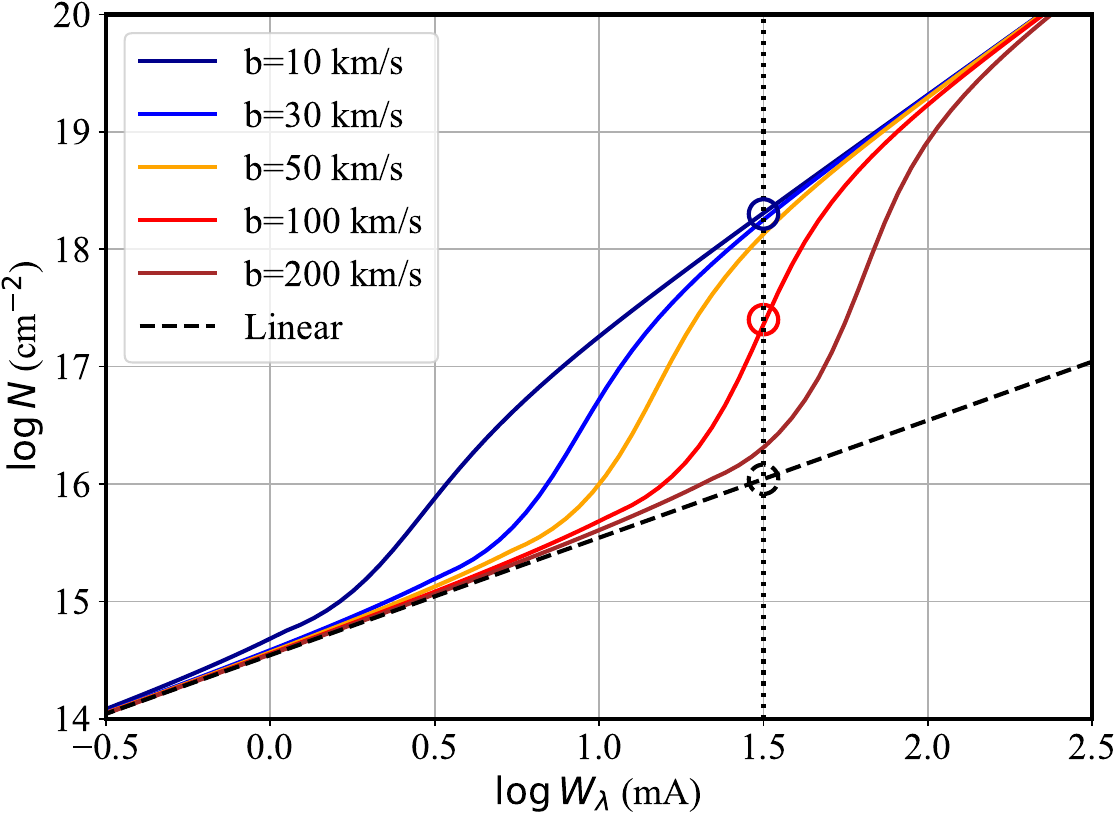}
     \includegraphics[width=3.4in]{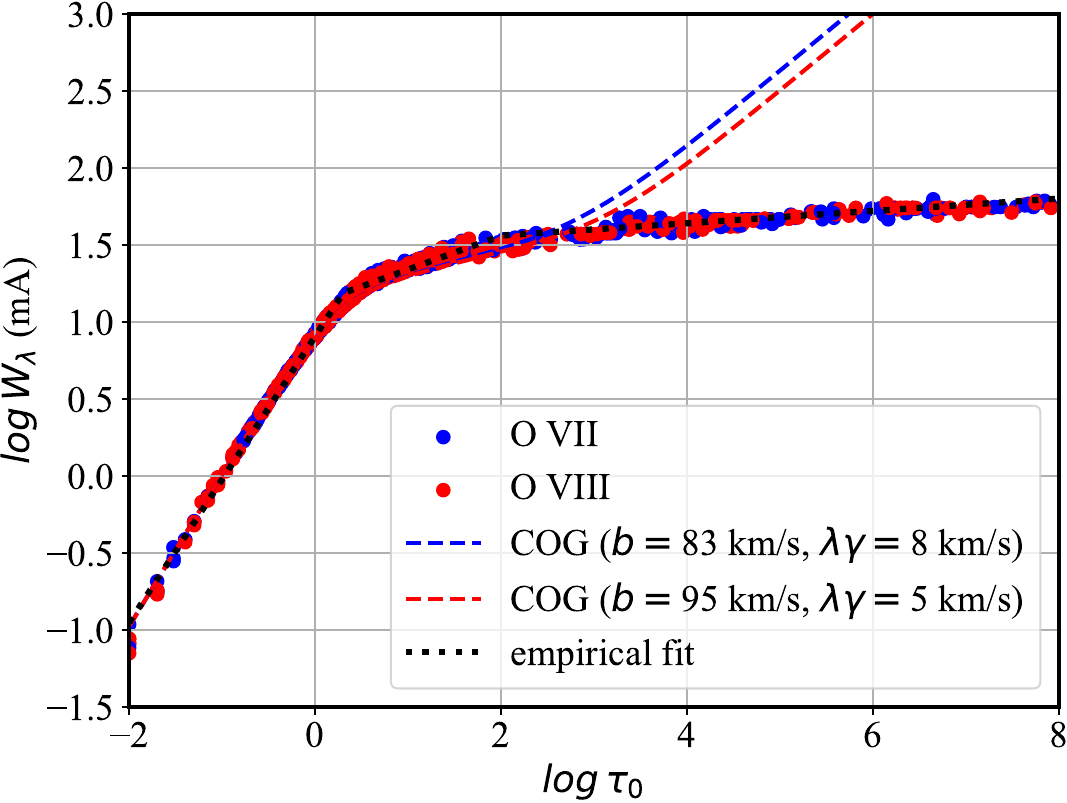}
    \caption{(Left:) Curves of growth (COG) for the resonance absorption line of \ovii\ at $\lambda_0=21.6$~\AA. The circles illustrates the
    column density that would be inferred from a putative measurement of 
    $W_{\lambda} \simeq 32$~m\AA, using respectively the linear approximation, or curves of growth for either $b=100$~km/s or $b=10$~\kms. (Right:) \texttt{SPEX}--measured values of the experimental $\tau_0$ parameter { and of the equivalent width $\Wl$} for our sample of \ovii\ and \oviii\ absorption lines, and comparison with the COG method.}
    \label{fig:COG}
\end{figure*}

 The velocity structure of a line plays an important role in the \cog\ analysis.
For a plasma in thermal equilibrium at a temperature $T$, the broadening parameter $b$ is given by
\begin{equation}
    b = \left(\dfrac{2 k T}{m}\right)^{\nicefrac{1}{2}} = 32.3 \left( \dfrac{T/10^6\, \text{K}}{m/16\, \text{amu}} \right)^{\nicefrac{1}{2}}\; \text{km s}^{-1}
    \label{eq:b}
\end{equation}
with $\sigma_v=b/\sqrt{2}$ the velocity dispersion. Given that \ovii\ and \oviii\ absorption lines 
are expected to occur at $\log T(\mathrm{K)=6-7}$, the thermal $b$ parameter is expected to be in the range
$30-100$~km~s$^{-1}$.

The curves of growth (COG) for the resonance \ovii\ line at $\lambda_0=21.6$~\AA\ are shown in the left panel of Fig~\ref{fig:COG},
as a function of the (unknown) velocity parameter $b$.
It is clear that large values of the equivalent width, which are often present in our
data in the form of upper limits to a non--detection, require a proper curve--of--growth analysis to
be converted to column densities. In particular, the linear approximation would severely underestimate the column density;  moreover, the unknown $b$ parameter of the line, which is a combination
of thermal and non--thermal broadening, cannot be estimated from the data, and therefore its
uncertainty must be taken into account in the estimate of the column density. Highlighted in the figure
is a nominal value of $\Wl\simeq32$~m\AA: the associated column density would be erroneously estimated
at $\log N \simeq 16$~cm$^{-2}$, if the proper COG is not taken into account, with an error that can be of more than one or two orders of magnitude. The figure highlights the importance of the $b$ parameter in the
estimation of column densities.

The right--hand panel of Fig.~\ref{fig:COG} shows the measured optical depth parameters and the corresponding equivalent widths as measured 
from the \spex\ fits, and the $\Wl(\tau_0)$ relationship according to \eqref{eq:COGRodgers} { (see App.~\ref{app:tables} for details)}. This
comparison shows that, in the linear and flat part of the COG, the measured $\Wl$ corresponds to
the values obtained from the COG method. In the damped part, which is only interesting to set
upper limits to poorly--constrained optical depths, the COG method yields a different result from the empirical method { discussed in App.~\ref{app:tables}}. 
This stems from the fact that, in that regime, the \spex\ parameter $\tau_0$ is no longer the optical depth at line center, as already noted in the foregoing, and therefore those curves are not used in
our application. An empirical fit with 
a step--wise linear function to
the $\Wl-\tau_0$ \spex\ measurements is also shown as a dotted line;  this is the relationship that
applies to our data and it can be used
to convert an upper limit to the $\tau_0$ parameter to an upper limit to the measured $\Wl$.

\subsection{Measurement of column densities}
\label{sec:colDen}
 The values of $\tau_0$ and of the equivalent widths are obtained from
the \spex\ fits presented in \paperOne, and they 
 are used to calculate the column densities using the \cog\ method
described in Sec.~\ref{sec:cog}, for a nominal value of $b=100$~\kms; the effect of this 
assumption is further examined in Sec.~\ref{sec:systematics}. Tables with the 
values of $\tau_0$, the equivalent width $\Wl$ and column densities for all systems in our sample are reported in Tables~\ref{tab:tauEWN-0} through~\ref{tab:tauEWN-7} in Appendix~\ref{app:tables}.

In addition to the measurements, Tables~\ref{tab:tauEWN-0} through~\ref{tab:tauEWN-7} also provide upper limits to these three quantities.
The upper limits to $\tau_0$ are calculated following the method described in Sec.~\ref{sec:ULMethods}, and they are reported for both the
"Feldman--Cousins" and "sensitivity" methods. These upper limits are then used to
calculate upper limits on $\Wl$ using the known experimental relationship between
$\tau_0$ and $\Wl$ (see right--hand panel of Fig.~\ref{fig:COG}).
Finally, the upper limits on $\Wl$ are used to calculate upper limits to the
column density using the COG equations of Sec.~\ref{sec:cog}. In so doing,
we use the statistical properties of the measured $\tau_0$ parameter (for which
we always have best--fit values and errors) to obtain
upper limits to the physically interesting quantities $\Wl$ and $N$ in a consistent manner.

\begin{figure*}
    \centering
    \includegraphics[width=\linewidth]{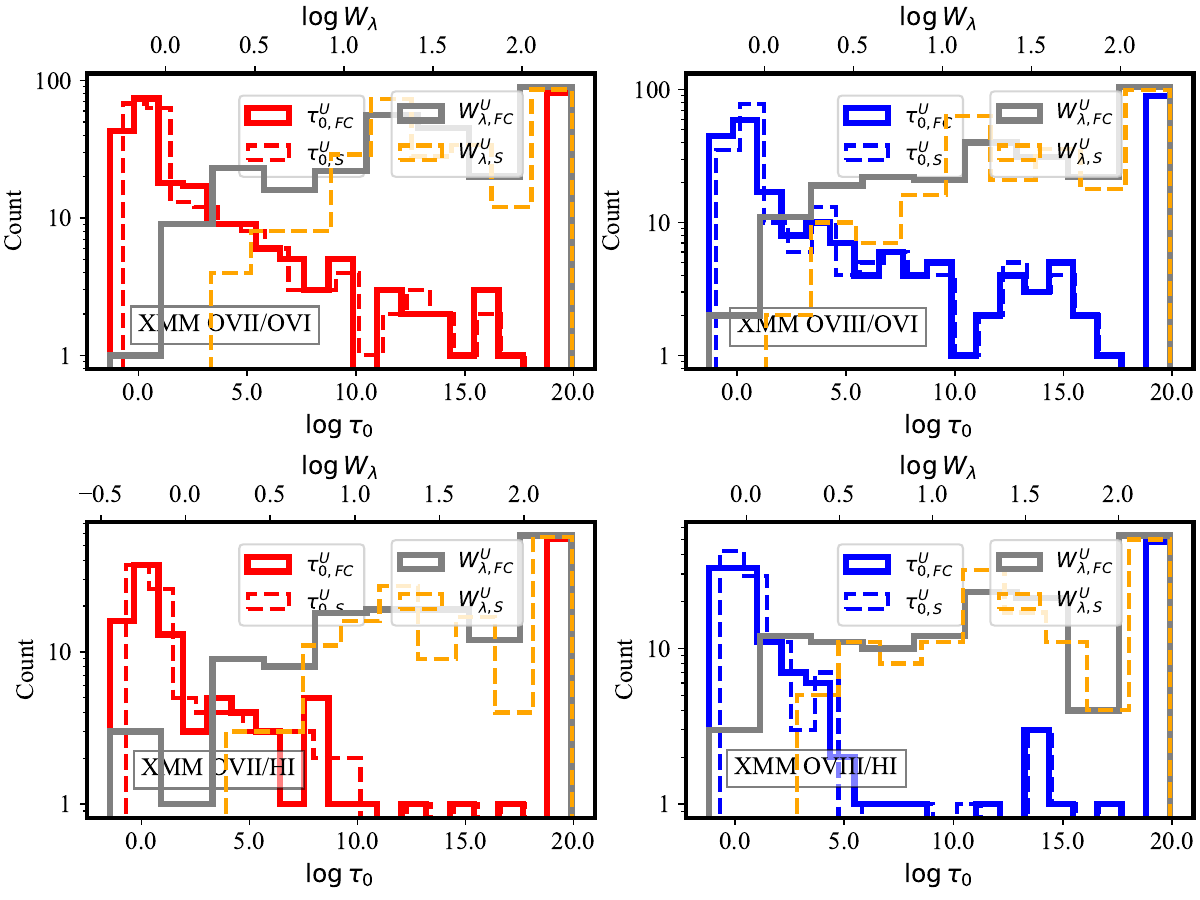}
    \caption{Distribution of upper limits to the $\tau_0$ parameter and to the equivalent width
    $\Wl$ of the lines for the \xmm\ fits.}
    \label{fig:tauEWULXMM}
\end{figure*}
 
\begin{figure*}
    \centering
    \includegraphics[width=\linewidth]{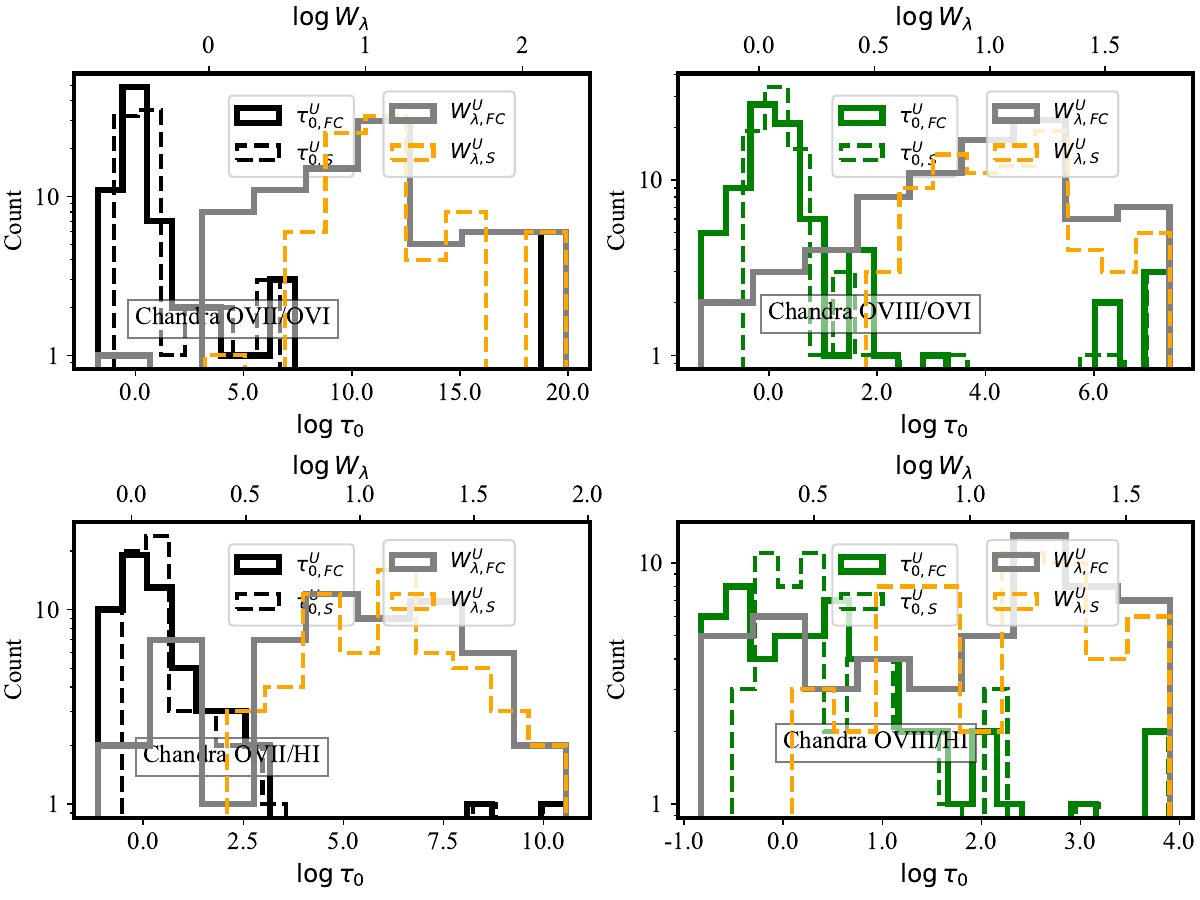}
    \caption{Distribution of upper limits to the $\tau_0$ parameter and to the equivalent width
    $\Wl$ of the lines for the \chandra\ data.}
    \label{fig:tauEWULChandra}
\end{figure*}

Figures~\ref{fig:tauEWULXMM} and \ref{fig:tauEWULChandra} display the distributions
of the upper limits to the $\tau_0$ parameters and to the corresponding equivalent widths,
showing that a significant number of fits
have a poorly constrained optical depth parameter, as indicated by the long tails of the distributions. In fact, many fits have a virtually unconstrained $\tau_0$ parameter 
(e.g., fits with $\log \tau_0 \simeq 20$ in Fig.~\ref{fig:tauEWULXMM}), which is the result 
of the low counts in several of the spectra. Accordingly, the corresponding equivalent widths also 
have broad distributions that reach $\Wl \simeq 100$~m\AA, consistent with 
the empirical relationship in the right panel of Fig.~\ref{fig:COG}. Such large values of
the equivalent width are a result of the low counts and the resolution of the spectra.
Distributions of the upper limits to the column densities are further discussed in the following section.

\def\yes{Y}
\def\no{N}
\def\maybe{?}
\begin{table*}
    \centering
    \begin{tabular}{lllHHl|p{1.2cm}p{1cm}l|llp{0.6cm}p{0.6cm}p{0.65cm}p{0.65cm}| l}
    \hline
    \hline\\
    \footnotesize
    Source & \# & $z$ & & & $\Delta C$ & $\tau_0$ & $\Wl$ & $\log N$ &  $\tau_{0,FC}^U$ & $\tau_{0,S}^U$ &  $W_{\lambda,FC}^U$ &$W_{\lambda,S}^U$ &  $\log N_{FC}^U$ &$\log N_{S}^U$ & WHIM\\[5pt]
    \hline
    \multicolumn{15}{c}{XMM OVII/OVI}\\ 
pks0405 & 261 & 0.4089 & 110.2 & 86 & 7.0 & $1.00e+07$ $\pm^{1.00e+20}_{9.95e+06}$ & $56.03\pm^{28.78}_{9.43}$  &  $18.531\pm^{0.512}_{0.292}$  &8.1e+19 & 8.3e+19 & 1.9e+02 & 1.9e+02 & 19.8 & 19.8 & \maybe \\ 
1es1553 & 5 & 0.1878 & 102.3 & 88 & 7.5 &  $0.57\pm^{0.30}_{0.25}$ & $5.23\pm^{2.16}_{2.08}$  &  $15.329\pm^{0.181}_{0.248}$  &1.00 & 0.45 & 8.13 & 3.90 & 15.56 & 15.18 & \yes\\ 
1es1553 & 4 & 0.1876 & 102.1 & 88 & 7.8& $0.60\pm^{0.32}_{0.26}$ & $5.35\pm^{2.21}_{2.05}$  &  $15.340\pm^{0.182}_{0.238}$  &1.05 & 0.48 & 8.54 & 4.10 & 15.59 & 15.21 & \yes \\ 
pg1211 & 180 & 0.0512 & 45.2 & 34 & 13.5& $2.02e+02$ $\pm^{3.78e+05}_{1.84e+02}$ & $32.99\pm^{14.70}_{7.97}$  &  $17.468\pm^{0.810}_{0.720}$  &3.1e+05 & 3.1e+05 & 5.0e+01 & 5.0e+01 & 18.4 & 18.4 & \maybe \\ 
tons180 & 308 & 0.0456 & 39.9 & 45 & 14.4 & $6.94e+01$ $\pm^{4.72e+03}_{5.61e+01}$ & $30.09\pm^{8.67}_{6.73}$  &  $17.219\pm^{0.651}_{0.622}$  &3.9e+03 & 3.9e+03 & 4.2e+01 & 4.2e+01 & 18.1 & 18.1 & \yes\\ 
3c273 & 23 & 0.1466 & 91.0 & 79 & 17.9 & $0.81\pm^{0.31}_{0.24}$ & $6.89\pm^{1.78}_{1.78}$  &  $15.473\pm^{0.127}_{0.156}$  &1.25 & 0.45 & 10.00 & 3.90 & 15.68 & 15.18 & \yes\ \maybe\\ 
3c273 ($\star)$& 21 & 0.0902 & 100.6 & 47 & 78.3 & $2.95\pm^{0.87}_{0.66}$ & $15.43\pm^{1.88}_{1.99}$  &  $15.983\pm^{0.126}_{0.118}$  &4.17 & 1.26 & 18.18 & 10.09 & 16.17 & 15.69 & \yes \\ 
\hline
\multicolumn{15}{c}{Chandra OVII/OVI}\\ 
pks0405 & 45 & 0.1657 & 45.8 & 38 & 7.2 & $1.30e+05$ $\pm^{1.00e+20}_{6.50e+04}$ & $49.94\pm^{36.84}_{0.00}$  &  $18.356\pm^{0.713}_{0.000}$  &8.1e+19 & 8.3e+19 & 1.9e+02 & 1.9e+02 & 19.8 & 19.8 & \maybe \\ 
\hline
\multicolumn{15}{c}{XMM OVIII/OVI}\\ 
1es1028 & 3 & 0.3373 & 92.6 & 96 & 6.6 & $5.01\pm^{15.34}_{3.36}$ & $19.15\pm^{7.90}_{8.13}$  &  $16.749\pm^{0.868}_{0.651}$  &19.83 & 15.43 & 25.61 & 24.23 & 17.46 & 17.30 & \yes\\ 
pks2155 & 267 & 0.0571 & 76.8 & 59 & 7.3 & $0.25\pm^{0.26}_{0.08}$ & $2.47\pm^{2.23}_{0.68}$  &  $15.306\pm^{0.315}_{0.150}$  &0.52 & 0.28 & 4.39 & 2.49 & 15.59 & 15.31 & \yes \\ 
pks0405 & 259 & 0.3633 & 110.1 & 94 & 8.0 & $2.92e+04$ $\pm^{4.34e+09}_{2.88e+04}$ & $41.89\pm^{18.66}_{5.41}$  &  $18.724\pm^{0.551}_{0.283}$  &3.5e+09 & 3.6e+09 & 7.3e+01 & 7.3e+01 & 19.5 & 19.5 & \maybe  \\ 
1es1553 & 6 & 0.1898 & 25.6 & 27 & 8.2 & $0.98\pm^{0.60}_{0.48}$ & $7.80\pm^{2.97}_{3.21}$  &  $15.893\pm^{0.190}_{0.285}$  &1.85 & 0.89 & 14.44 & 7.30 & 16.33 & 15.86 & \maybe \\ 
mrk421 & 114 & 0.0101 & 95.4 & 57 & 9.7& $0.09\pm^{0.11}_{0.02}$ & $0.98\pm^{1.07}_{0.22}$  &  $14.884\pm^{0.335}_{0.113}$  &0.20 & 0.11 & 1.79 & 1.02 & 15.16 & 14.90 & \yes \\ 
1es1553 & 9 & 0.3787 & 113.1 & 88 & 15.3 & $1.03\pm^{0.44}_{0.35}$ & $8.17\pm^{2.35}_{2.25}$  &  $15.920\pm^{0.149}_{0.180}$  &1.68 & 0.65 & 13.14 & 5.46 & 16.24 & 15.70 & \yes\ \maybe \\ 
1es1553 & 8 & 0.3113 & 113.1 & 91 & 17.1 & $1.01\pm^{0.37}_{0.33}$ & $8.36\pm^{2.16}_{2.30}$  &  $15.934\pm^{0.136}_{0.181}$  &1.59 & 0.58 & 12.52 & 4.88 & 16.19 & 15.64 & \yes\ \maybe\\ 
ngc7469 & 121 & 0.0096 & 84.3 & 74 & 45.8 & $2.35\pm^{0.90}_{0.64}$ & $14.20\pm^{2.10}_{2.39}$  &  $16.313\pm^{0.171}_{0.166}$  &3.58 & 1.27 & 17.57 & 10.16 & 16.60 & 16.05 & \no\\ 
ngc7469 & 123 & 0.0115 & 85.8 & 74 & 53.4 & $3.35\pm^{1.86}_{1.00}$ & $16.68\pm^{2.47}_{2.48}$  &  $16.516\pm^{0.232}_{0.204}$  &5.63 & 2.36 & 19.42 & 16.03 & 16.78 & 16.46 & \no\\ 
ngc7469 & 122 & 0.0099 & 94.8 & 74 & 53.8 & $2.92\pm^{1.08}_{0.81}$ & $15.57\pm^{2.31}_{2.32}$  &  $16.421\pm^{0.204}_{0.178}$  &4.49 & 1.56 & 18.47 & 12.29 & 16.68 & 16.18 & \no \\ 
\hline
\multicolumn{15}{c}{Chandra OVIII/OVI}\\ 
3c273 & 77 & 0.0902 & 29.9 & 38 & 6.6& $3.34\pm^{4.82}_{2.01}$ & $16.30\pm^{6.20}_{6.48}$  &  $16.483\pm^{0.620}_{0.453}$  &8.97 & 5.63 & 21.51 & 19.42 & 16.99 & 16.78 & \yes\\ 
\hline
\multicolumn{15}{c}{XMM OVII/HI}\\ 
pg0804 & 52 & 0.0502 & 46.3 & 37 & 7.4& $3.75e+05$ $\pm^{1.00e+20}_{3.75e+05}$ & $51.10\pm^{22.76}_{nan}$  &  $18.393\pm^{0.495}_{nan}$  &8.1e+19 & 8.3e+19 & 1.9e+02 & 1.9e+02 & 19.8 & 19.8 & \maybe \\ 
s50716 & 159 & 0.0883 & 48.9 & 39 & 8.0 & $4.84e+04$ $\pm^{5.27e+07}_{4.84e+04}$ & $46.61\pm^{9.43}_{nan}$  &  $18.239\pm^{0.292}_{nan}$  &4.3e+07 & 4.4e+07 & 6.1e+01 & 6.2e+01 & 18.7 & 18.7 & \maybe \\ 
pg1116 & 75 & 0.0838 & 46.6 & 32 & 8.2&$2.02e+07$ $\pm^{1.00e+20}_{1.99e+07}$ & $56.03\pm^{26.85}_{9.43}$  &  $18.531\pm^{0.487}_{0.292}$  &8.1e+19 & 8.3e+19 & 1.9e+02 & 1.9e+02 & 19.8 & 19.8 & \maybe \\ 
\hline
\multicolumn{15}{c}{Chandra OVII/HI}\\ 
mr2251 & 28 & 0.0633 & 50.5 & 39 & 6.8 & $1.41e+01$ $\pm^{1.71e+03}_{1.13e+01}$ & $26.21\pm^{11.67}_{10.78}$  &  $16.858\pm^{0.959}_{0.875}$  &1.4e+03 & 1.4e+03 & 4.1e+01 & 4.1e+01 & 18.0 & 18.0 & \no\\ 
h1821 & 16 & 0.1982 & 30.7 & 38 & 6.9 & $2.58\pm^{2.94}_{1.42}$ & $14.40\pm^{5.03}_{5.52}$  &  $15.920\pm^{0.346}_{0.307}$  &6.17 & 3.60 & 19.81 & 17.59 & 16.30 & 16.13 & \yes \\ 
pg1116 & 38 & 0.1337 & 41.7 & 34 & 7.0 & $2.09e+01$ $\pm^{5.20e+02}_{1.85e+01}$ & $26.21\pm^{12.56}_{10.05}$  &  $16.858\pm^{1.012}_{0.828}$  &4.6e+02 & 4.4e+02 & 3.9e+01 & 3.9e+01 & 17.9 & 17.9 & \yes \\ 
mr2251 & 29 & 0.0638 & 48.0 & 38 & 8.0 & $1.67e+01$ $\pm^{8.98e+01}_{1.31e+01}$ & $26.82\pm^{7.73}_{9.10}$  &  $16.915\pm^{0.674}_{0.777}$  &9.9e+01 & 8.5e+01 & 3.7e+01 & 3.5e+01 & 17.7 & 17.6 & \no\\ 
\hline
\multicolumn{15}{c}{XMM OVIII/HI}\\ 
pks2155 & 129 & 0.0571 & 51.9 & 51 & 7.0 & $0.29\pm^{0.18}_{0.11}$ & $2.77\pm^{1.52}_{1.06}$  &  $15.360\pm^{0.213}_{0.226}$  &0.53 & 0.24 & 4.48 & 2.15 & 15.60 & 15.24 & \yes \\ 
pks0405 & 121 & 0.1946 & 47.4 & 42 & 13.4 & $6.59e+06$ $\pm^{1.00e+20}_{6.26e+06}$ & $55.23\pm^{17.58}_{4.86}$  &  $19.155\pm^{0.338}_{0.130}$  &8.1e+19 & 8.3e+19 & 1.9e+02 & 1.9e+02 & 20.4 & 20.4 & \maybe\\ 
mr2251 & 33 & 0.0619 & 63.0 & 47 & 16.5& $3.43\pm^{2.87}_{1.58}$ & $16.68\pm^{4.32}_{4.60}$  &  $16.516\pm^{0.423}_{0.353}$  &6.97 & 3.67 & 20.35 & 17.67 & 16.87 & 16.61 & \no \\ 
mr2251 & 34 & 0.0628 & 50.9 & 49 & 38.9& $2.40e+01$ $\pm^{4.44e+01}_{1.52e+01}$ & $26.43\pm^{3.23}_{4.45}$  &  $17.548\pm^{0.344}_{0.502}$  &7.2e+01 & 4.9e+01 & 3.4e+01 & 3.1e+01 & 18.3 & 18.1 & \no\\ 
mr2251 & 36 & 0.0638 & 45.7 & 49 & 45.5& $1.98e+01$ $\pm^{3.19e+02}_{1.14e+01}$ & $27.05\pm^{4.73}_{4.03}$  &  $17.617\pm^{0.474}_{0.455}$  &2.8e+02 & 2.7e+02 & 3.8e+01 & 3.8e+01 & 18.5 & 18.5 & \no \\ 
mr2251 & 35 & 0.0633 & 44.9 & 49 & 45.7& $3.03e+01$ $\pm^{1.10e+02}_{1.96e+01}$ & $27.68\pm^{2.67}_{4.12}$  &  $17.686\pm^{0.274}_{0.464}$  &1.3e+02 & 1.1e+02 & 3.7e+01 & 3.7e+01 & 18.5 & 18.5 & \no  \\ 
ngc7469 & 46 & 0.0098 & 85.7 & 65 & 51.6 & $2.83\pm^{1.01}_{0.77}$ & $15.21\pm^{2.25}_{2.26}$  &  $16.392\pm^{0.195}_{0.170}$  &4.32 & 1.47 & 18.31 & 11.62 & 16.67 & 16.13 & \no \\ 
\hline
\multicolumn{15}{c}{Chandra OVIII/HI}\\ 
h1821 & 13 & 0.0678 & 39.4 & 38 & 7.8 & $2.02\pm^{1.67}_{0.98}$ & $12.65\pm^{4.41}_{4.48}$  &  $16.202\pm^{0.350}_{0.282}$  &4.15 & 2.19 & 18.15 & 15.77 & 16.65 & 16.44 & \yes\\ 
    \hline
    \hline
    \end{tabular}
    \caption{Systems with possible absorption line detections with $\Delta C \geq 6.6$, positive $\tau_0$ parameter, and a \gof\ \cmin/\dof$\leq 2$. A WHIM categorization of "Y" means that the line is likely of genuine WHIM origin, a question mark means that the line is of uncertain nature, and an "N" indicates that the line is unlikely of WHIM origin. See Sec.~\ref{sec:cosmo-detections} for criteria of this classification. (The first 4 column of this table are from 
    Table~13 of \paperOne; columns 5--12 are also available for all sources in Tables~\ref{tab:tauEWN-0}--\ref{tab:tauEWN-7} in this paper.)}
    \label{tab:detections}
\end{table*}

Results for the \nDet\ systems with a possible detection of absorption lines are also summarized in Table~\ref{tab:detections}. As discussed in \paperOne, we also include one system 
marked with a star (for the source 3C273) which
features a reduced \cmin\ slightly above the arbitrary threshold of 2.0. Hereafter we will
refer to these 33+1 possible detections  as the baseline sample of systems with possible
absorption lines. 
The values reported in this table illustrate that some of the possible absorption--line detections
have poor constraints on the equivalent width and therefore column density of the line.
For example, the first system is a possible \ovii\ absorption line in the \xmm\ data of PKS~0405-123, at 
the fixed redshift of an \ovi\ absorption line. The spectrum, which was shown in Figure~5 of 
\paperOne, has a rather
low count rate of $\sim (2-4)\times10^{-3}$~counts~s$^{-1}$~\AA$^{-1}$ at the position of the line (near 30.4~\AA),
with several datapoints consistent with zero counts at those wavelengths.  Such low count rates
contribute to the poor constraints in the physical parameters of interest, and raise
concerns as to reality of this possible absorption feature. 
A similar situation
applies to other sources that have poor constraints in Table~\ref{tab:detections}. A comprehensive 
assessment
of the presence of X--ray absorption in these low-count systems is discussed in Sec.~\ref{sec:cosmo-detections}.

\section{Cosmological baryon density of the X--ray absorbers}
\label{sec:cosmo}

The large number of sources and
redshift path covered by this sample can be used to 
constrain the cosmological density of baryons implied by these X--ray measurements.
 We refer to this density as \ObX, to imply that it only comprises
the hotter portion of baryons as tracked by \ovii\ and \oviii\ X--ray absorption lines. 
In this section we review the methods of analysis that lead to the cosmological 
constraints from the X--ray data, using either positive detections
or upper limits. 

\subsection{Method of analysis to constrain the baryon density}
\label{sec:cosmo-method}
The method makes use of the measured column densities from the X--ray data and, for
systems with upper limits, it is
complemented by the analysis of \eagle\ simulations \citep{wijers2019} to predict
the amount of \ovii\ and \oviii\ column density expected from the detected FUV systems.
The method  was
presented in a previous study \citep{spence2023}, where it was applied to the analysis of
one of the brightest sources analyzed in this paper, viz. \es. In that analysis it was
found that the non--detection of absorption lines could be interpreted with an upper limit $\ObXbEq \leq 0.90$ for reasonable choices of the temperature and abundance of the 
probed gas, meaning that the X--ray absorbing WHIM is consistent with being the repository of the missing baryons. Those preliminary results are now extended to all the sources in our sample.

\subsubsection{Method for detections}
The method of analysis for a system with a positive detection starts with the usual relationships
\begin{equation}
\begin{cases}
 \Omega_{\mathrm{WHIM,ion}} =\Omega_{\ion} \dfrac{\mu_H \, m_H}{m_{\ion}} \dfrac{1}{A \cdot f_{\ion}(T)}\\[10pt]
\label{eq:Omega}
        \Omega_{\text{ion}}= \dfrac{\rho_{\text{ion}}}{\rho_c} \text{, with }
\rho_{\ion} = m_{\ion} \dfrac{N_{\ion}}{D_i}.
\end{cases}
\end{equation}
These equations link the observable column density $N_{\ion}$ to the cosmological density
of baryons, with $D_i$ the distance to the source, $A$ the metal abundance of the gas, $f_{\ion}(T)$
the ionization fraction for the assumed temperature $T$, with all other
constants having the usual meaning (see \citealt{spence2023}). Equation~\ref{eq:Omega} applies to a single source
and a single ion/absorption line. 

The equation can 
 be generalized to multiple sources via
\begin{equation}
\OWHIMX= \dfrac{\mu_{ H} \, m_H}{\rho_c} \dfrac{{\sum_i} N_{H,i}}{\sum_i D_i}
\label{eq:OmegaWHIMX}
    \end{equation}
    \citep[see, e.g.,][]{schaye2001,nicastro2018}
where the sum extends over all available sources, and the $i$--th source probes
a cosmological distance $D_i$ with a detection of a total WHIM column density $N_{H,i}$, representing
the entire baryonic content of the WHIM. It is clear that care must be exercised in not double--counting absorbers that may
be probing the same $N_{H,i}$ column density (e.g., \ovii\ and \oviii\ that may be originating from the same gas phase); this
issue will be addressed in Sec.~\ref{sec:systematics} below.

\subsubsection{Method for upper limits}
In the case of upper limits to the detection of $N_{\ion}$, and therefore of $N_{H,i}$, it is not accurate to just
replace the measurements with upper limits of those quantities in \eqref{eq:Omega} or \eqref{eq:OmegaWHIMX}. 
In fact, upper limits are significantly affected by the limited resolution of the data, which means that 
the data are only sensitive to a small fraction of the possible WHIM present along the sightline.
As an example, consider entries 8 or 9 in Table~\ref{tab:tauEWN-1}, with upper limits of $\log N^U (\mathrm cm^{-2})\sim 20$. Such large values
simply mean that the the data cannot effectively constrain the equivalent width of the line,
and these values largely overestimate any reasonable column densities along those sightlines \citep[e.g., according to the \eagle\ simulations of][]{wijers2019}.

Instead, in \cite{spence2023} we proposed a method that consists of calculating the \emph{expectation} of the 
column density that is implied by the non--detection along the $i$--th source, viz. 
\begin{equation}
    \E[\nioni^+]=\sum_{j=1}^{n_i} \E[\nionj^+],
    \label{eq:nionMinus}
\end{equation}
which represents an \emph{average upper limit} to the column density of
the ion that was systematically not detected in a given source, with $n_i$ the number of absorption lines that were probed for
the $i$--th source (note that the index $j$ refers to the $j$--th absorption system for $i$--the source).
The expectation is intended as a probabilistic expectation based on the distribution of values for the 
ion $N_{\ion,j}$ \emph{given} the amount of detected FUV absorption (e.g., either \ovi\ or \hi) as calculated
according to the \eagle\ simulations \citep{wijers2019}. These expectations are 
a function of the
measured upper limit \NUL, with a higher upper limit naturally corresponding to a larger expectation,
and further take into account the measured FUV column density, given the positive
correlation between FUV (\ovi\ and \hi) and X--ray (\ovii\ and \oviii) column densities.  A complete description of this method is presented in Sec.~5 of \cite{spence2023}, which we follow in this paper.~\footnote{A more 
appropriate notation for the expectations in the right--hand side of \eqref{eq:nionMinus} would be
$\E[\nionj^+] \coloneq \E[\nionj\ | N_{\mathrm{FUV},j}, \NULeq]$, i.e., the expected ion column \emph{given} the measured value of
the FUV column density, and given the measured upper limit to the non--detection of the X--ray ion.}

In practice, with the observational upper limit \NUL obtained for each absorption line system of each source
(Tables~\ref{tab:tauEWN-0} through \ref{tab:tauEWN-7}), we 
calculate each expectation $\E[\nionj^+]$, and then sum according to \eqref{eq:nionMinus}. 
That number is used to calculate
the expectation on the column density of the associated hydrogen according to the usual
\begin{equation}
    \E[N_{H,i}]= \dfrac{\mu_H m_H}{m_{\ion} A f_{\ion}(T)}\E[\nioni^+],
\end{equation}
and from this the upper limit on $\OWHIMX$ is obtained from \eqref{eq:OmegaWHIMX}, viz.
\begin{equation}
    \ObXEq \leq  \dfrac{\mu_H \, m_H}{\rho_c\, A\, 
    f_{\ion}(T)} \dfrac{\sum_i\E[\nioni^+]}{\sum_i D_i}.
    \label{eq:OmegaWHIMXUL}
\end{equation}
with the sum at the numerator being the sum of the expected ion column
densities over all sources and all absorption line systems. 

In summary, the key reason for the use of such theory--based expectations, instead of simple data--based upper limits,
is that the resolution of the available X--ray data are quite limited, and 
theoretical predictions thus help in 
reducing the uncertainties associated with the X--ray data alone. In principle, the use of
different predictions based on different simulations would affect the results according to \eqref{eq:OmegaWHIMXUL}. Systematics associated with the use of \eagle\ results are further discussed in Sec.~\ref{sec:systematics}.

\subsection{Identification of possible WHIM X--ray absorption lines}
\label{sec:cosmo-detections}

The statistical significance of the  absorption lines in this sample
was already discussed in \paperOne. Here we summarize the main points.
The first consideration is the fact that each FUV prior sets a single
opportunity for the detection of X--ray absorption, without \emph{any} adjustments to
optimize the significance of detection of a given line.
This means that there is no need to use the 
framework of `redshift trials' \citep[see, e.g.][]{kaastra2006, nicastro2013, bonamente2019}, which would be needed for blind searches \emph{\`{a} l\`{a}} \cite{nicastro2018} or \cite{gatuzz2023} 
where the location of possible lines is unknown a priori. We therefore
take the view that each possible detection in Table~\ref{tab:detections} has a nominal
significance of detection equal to that indicated by the $\Delta C \geq 6.6$ statistic, which is 
$\geq 99$~\% for each of the 33+1 possible absorption lines. Second, { in Sec.~3.2 of \paperOne} we have used a simple
probabilistic model, based on the binomial distribution, to further show that 
such number of possible absorption-line detections
at this level of significance are highly unlikely to occur by chance, out of the \nsystems\ systems
investigated, when the sample of 33+1 possible detections is considered as a whole. The analysis 
presented in \paperOne\ therefore indicates that there is positive
evidence for the presence of absorption lines in this X--ray sample.

{ At any level of significance,  however, there is always a non--zero probability that any given detection is 
 spurious. In our data, one would expect $\sim 12$ of the \nsystems\ sources to randomly feature a detection statistic
that corresponds to $p=0.01$, and about $3-4$ to feature $p=0.003$ (i.e., the $p$--value that corresponds to a ``$3-\sigma$ detection''). Instead of raising the (arbitrary)  threshold for ``detection'', we proceed to further screen these possible detections via a number of physically and statistically  motivated criteria that are described below.}
The $\Delta C \geq 6.6$ criterion for a possible absorption--line detection {used in \paperOne} is
{therefore}
simply an initial necessary condition 
for further consideration as a possible WHIM absorption--line system. 
In particular, in Sec.~\ref{sec:DeltaC} below we discuss the possibility of inaccurate  detection results
based on the $\Delta C$ statistic for certain low-count spectra. {We also address the effect of
the possible presence of spurious
detections on the determination of cosmological parameters in Sec.~\ref{sec:sysDetection} by raising
the threshold of $p$ values required for consideration as a possible detection.}

 \subsubsection{Confusion with Galactic or intrinsic lines}
 \label{sec:confusion}
 First condition for further consideration as a possible WHIM  absorption lines is that
 the redshift of the absorption line cannot be too close to either $z=0$ or $z=z_{\mathrm QSO}$,
 in order to be considered of genuine intergalactic origin. 
 
 According to this criterion, we note that the
systems in MR2251 are all at a redshift that is near the redshift of the quasar $z_{\mathrm QSO}=0.064$,
and system 46 for NGC~7469 is likely of Galactic origin, given that the redshift implies a peculiar velocity of $3,000$~\kms\ from $z=0$. The redshift $z=0.147$ for system 23 of 3C273 also implies a peculiar velocity of $\sim -3,000$\kms\ relative to  $z_{\mathrm QSO}=0.158$. Therefore, an intrinsic origin cannot be completely ruled out for these systems.

\subsubsection{Confusion with other Galactic lines}
\label{sec:confusionGalactic}
In Sec.~2.5.3 of \paperOne\ we discussed several lines in the 21.6-23.5~\AA\ band that 
can originate from Galactic \oi--\ovi\ \citep[e.g., using the atomic data of][]{gatuzz2015}. As a result,
we require that the wavelength of the absorption line not be too close to that of other known Galactic lines.

The $z=0.0902$ redshift for system 21 in 3C273 (marked with a star in the table) is such that
its \ovii\ K-$\alpha$ lines is at $\lambda=23.548$, which is close to that of Galactic \oi\ at $\lambda=23.506$, as already noted in Sec.~3.4.1 of \paperOne\ and in \cite{ahoranta2020}. The wavelength difference correspond to two
bins in the \xmmshort\ spectra, and therefore an \oi\ origin cannot be completely ruled out. The fact that
the associated \oviii\ line (system 77) is also significant, however, suggests a genuine WHIM origin. 

\subsubsection{Absorption by the circum--galactic medium of galaxies}
A plausible explanation for these absorption line features is also  absorption
by the warm--hot circum--galactic medium (often referred to as WCGM or CGM) of line--of--sight
galaxies \citep[e.g.][]{tumlison2017, das2019, wijers2022,  tuominen2023}.
We therefore require that there are no known massive galaxies near the absorber
as a condition for consideration as a possible WHIM absorption line.

For this purpose, in \paperOne\ we conducted a search for known galaxies within 1~Mpc in the plane of the sky, and with a peculiar  line--of--sight velocity of $\pm 1,000$~\kms, corresponding 
to approximately $\Delta z =\pm 0.0034$ of the absorber's nominal redshift. System 45 of PKS~0405  had a galaxy at a projected distance of 0.12~Mpc, indicating possible contamination by a circum--galactic medium.
System 28 of MR~2251 (\oviii/\ovi) and systems 33--36 (\oviii/\hi) had a galaxy at an impact parameter of just 50~kpc, indicating a likely circum-galactic origin for the absorption; those system was also considered to be likely associated to the emitting source according to the criterion in Sec.~\ref{sec:confusion} above. These systems  are therefore
unlikely to be genuine WHIM systems. Other systems, including those towards Ton~S180 and \es\ were previously shown to be unlikey of CGM origin based on the impact parameter and mass
of the intervening galaxies \citep{ahoranta2021,spence2023}.

\subsubsection{Reliability of the $\Delta C$ statistic for certain low--count sources}
\label{sec:DeltaC}
\begin{figure*}
    \centering
    \includegraphics[width=3.4in]{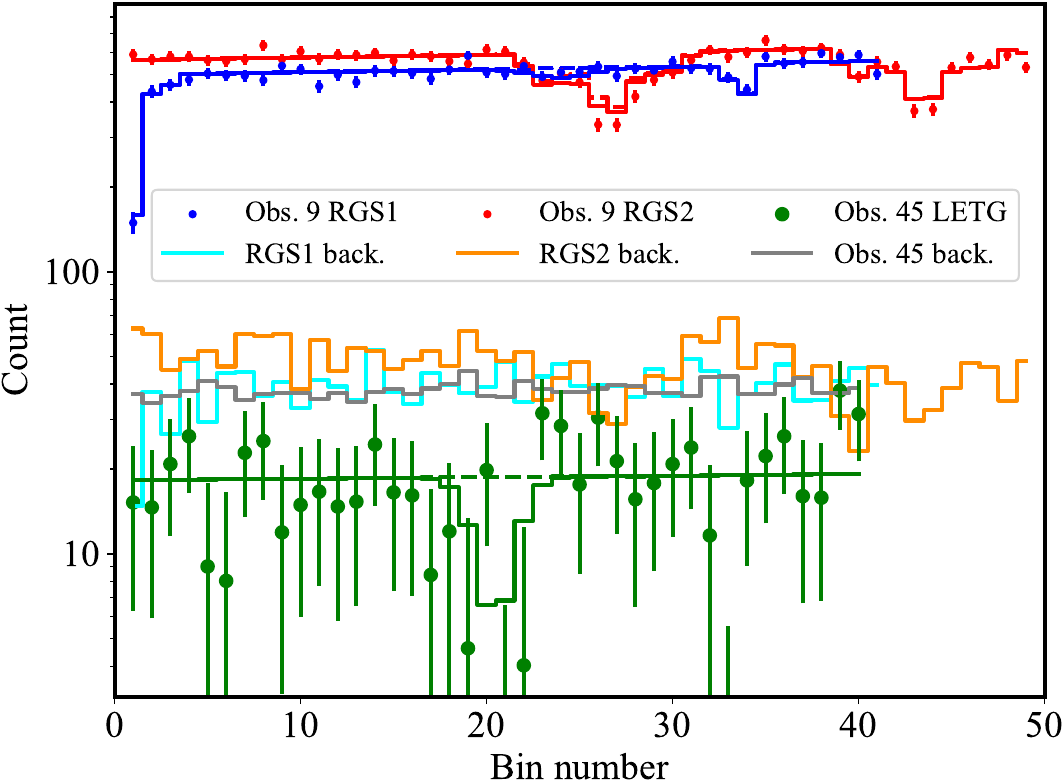}
    \includegraphics[width=3.2in]{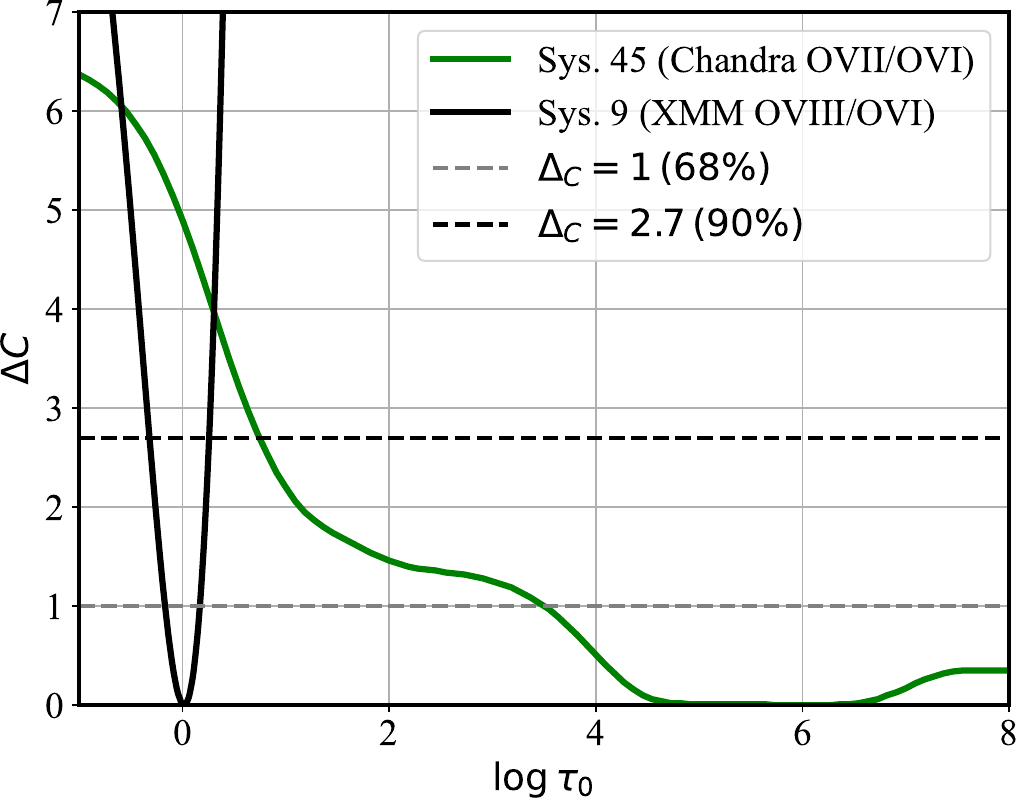}
    \caption{Illustration of the effects of low-count spectra on the $\Delta C$ statistic. (Left:) High--count RGS1 and RGS2 \xmm\ spectra for system 9 (\oviii/\ovi) and low--count HRC/LETG \chandra\ spectrum for system 45 (\ovii/\ovi). Spectra are in bin and counts units, so different wavelengths apply to different spectra. (Right:) Behavior of the $\Delta C$ statistic as a function of the model parameter $\tau_0$ 
    for the two observations.}
    \label{fig:45}
\end{figure*}

The results of Table~\ref{tab:detections} show that certain spectra with $\Delta C \geq 6.6$ have poorly constrained parameters for the \line\ component. For this purpose, in the left panel of 
Fig.~\ref{fig:45} we show the spectra of two representative cases:  \es\ system 9 (\oviii/\ovi, with 2~Ms of \xmm\ exposure)
with tight constraints on the model component; and PKS~0405 system 45 (\ovii/\ovi, with 380~ks of \chandra\ exposure) with
much looser constraints. The former has several hundred counts per bin \citep[see, e.g.][]{nicastro2018, spence2023} and the source is
significantly above the background, while the latter has only few counts per bin with a background that exceeds the source count rate. 

In \spex, the \cmin\ and therefore $\Delta C$ statistics are duly calculated
from the total integer--valued count spectrum (in the figure, the spectra are background--subtracted). The calculation, however, assumes a fixed background, as an approximation that ignores the background variability. It is useful to point out that in \texttt{XSPEC} such case of
Poisson source with Poisson background would be handled with a modification of the \cstat\ that is referred to
as the $W$--stat \citep{AppendixB}, which is however known to have problems in the low--count regime \citep{vianello2018}. The right panel of Fig.~\ref{fig:45} shows the behavior of the $\Delta C$ 
statistic as a function of the logarithm of the parameter $\tau_0$. For the high--count spectrum 9,
where incidentally the large mean results in a good approximation of the Poisson distribution with a normal
distribution, the $\Delta C$ distribution follows approximately a parabolic curve, as is expected of normal data or in the asymptotic extensive (i.e., large number of data points) limit 
\citep[see, e.g., Chapter~9 of][]{james2006}.~\footnote{The $\Delta C$ statistic is a 
logarithmic likelihood--ratio statistic, and its distribution as a function of the parameter $\tau_0$ in the neighborhood
of the best--fit value is expected to be quadratic, under certain regularity conditions {\citep[e.g.][]{wilks1938, wilks1962, james2006}.}} For the
low--count spectrum 45, instead, the $\Delta C$ curve has a more complex shape. Moreover its
lowest value of $\Delta C=0$ (i.e., the point of minimum of the $C$ statistic) is not well determined.
Such shape of the $\Delta C$ curve is associated with the low--count nature of the regression,
and likely also with the assumptions of a fixed background and the fact that the spectrum is background--dominated. Similar $\Delta C$ curves apply to the other background--dominated low--count spectra in 
Table~\ref{tab:detections}.

In practice, this kind of `problematic'  $\Delta C$ curves 
mean that the best--fit value and the confidence intervals 
of $\tau_0$ (and therefore of $\Wl$ and $N$) may be inaccurate. This is illustrated in the
figure: the use of a value of $\Delta C=2.6$
as a criterion for a 90\% confidence interval would give a 
 confidence interval on $\tau_0$ that is \emph{orders of magnitude}
larger than with the default value of $\Delta C=1$ (a formal 68\% confidence interval), whereas for 
spectrum 9 the two 68\% and 90\% confidence intervals only differ by a small factor, as expected for
a quadratic log--likelihood curve. 
Accordingly, we opt to consider these possible detections as uncertain, and further discuss the impact
of this choice on the cosmological results in Sec.~\ref{sec:systematics}. 
For consistency in the analysis of all the sources and for the sake of completeness, in Tables~\ref{tab:tauEWN-0} through
\ref{tab:tauEWN-7} and in Table~\ref{tab:detections} we chose to retain the 
confidence intervals that appear suspiciously large. 

{ We also note that possible detections with this kind of non--normal likelihood profiles
correspond to the largest estimated column densities in Table~\ref{tab:detections}. 
For a simple geometry in which the sightline intercepts a length $L$ of a constant--density filament, the ion column density $N$ can be used to estimate the filament length via
\begin{equation}
    L = 4.9 \left(\dfrac{N}{10^{15} \text{cm$^{-2}$}}\right) \left(\dfrac{A}{0.1\text{ Solar}}\right)^{-1} \left(\dfrac{f_{\text{ion}}}{0.1}\right)^{-1} \left( \dfrac{\delta_b}{30}\right)^{-1} \text{Mpc}
\end{equation}
where $\delta_b$ is the baryon overdensity in the WHIM filament for a standard $\Lambda$CDM cosmology \citep{planck2016-cosmology}, $f_{\text{ion}}$ the usual ion fraction, and $A$ the 
elemental abundance relative to Solar \citep{anders1989}. Even amidst substantial uncertainties in some of these 
parameters, it is clear that values of  order $\log N \simeq 18$ are not tenable,
as they would correspond to filament lengths that are substantially longer than what is typically seen in simulations
\citep[e.g.][]{wijers2019,tuominen2021,holt2022}. These estimates further
justify their screening from further consideration for the WHIM baryon budget.
}

\subsubsection{Results of the selection}
\label{sec:selection}
As a result of the aforementioned criteria, in Table~\ref{tab:detections} we have identified 7 \ovii\ systems and 8 \oviii\ systems
as "likely" X--ray WHIM absorption line{s}.
This is the baseline sample that will
be used for cosmological inferences on the WHIM X--ray baryon density.

Of these 15 possible detections, and excluding the
2C273 system which was identified as a poor fit, the largest \gof\ statistic for the
detection of an absorption line is $\Delta C= 17.9$. This statistic has a formal
$p$--value of $2.3\times 10^{-5}$ according to its asymptotic
distribution according to the null--hypothesis, as discussed in \paperOne. 
This $p$-value also corresponds the probability of exceeding  
  values $\pm 4.2$ in a standard normal distribution, which is commonly referred to as a 
  "4.2--$\sigma$  detection" in the astronomical community. 
  It is necessary to point out that the normal
  distribution plays no role in the analysis of these count data, or in the statistical
  interpretation of the \gof\ values; it is only used as a comparison, given the familiarity of many readers with
  the normal distribution. The threshold for consideration of a possible absorption line
  features as a genuine detection was $\Delta C=6.6$, which has a formal $p$--value of 0.01
  that likewise
  corresponds to a "2.7--$\sigma$ detection". 

   It is therefore clear that the X--ray absorption line detections presented in
  this work are of limited statistical quality, especially when compared to the
  FUV absorption--line detections used as priors, which have much larger significance of detection
  \citep[see][]{danforth2016, tilton2012}.  Sources of systematic error
  associated with the analysis of the data  have the  potential to further
  reduce the significance of some of the X--ray detections, as discussed in \paperOne. In Sec.~\ref{sec:systematics}
  of this paper, we address additional sources of systematic errors associated with the
  cosmological interpretation of the results, including the possibility that some of these
  likely detections are spurious { and by raising the threshold to a $p$-value that corresponds to a ``$3-\sigma$ detection''}.

\subsection{Constraints on \ObX}
\label{sec:results}
We present two sets of constraints on the cosmological density
of X--ray WHIM baryons. First, we use the 7 \ovii\ and 8 \oviii\ likely WHIM
X--ray absorption lines in Sec.~\ref{sec:cosmo-detections} to obtain an estimate of \ObX\
according to \eqref{eq:OmegaWHIMX}. The cosmological results from this analysis therefore
assumes that the possible absorption lines marked with a "Y" in Table~\ref{tab:detections}
are indeed real, and associated with the extragalactic WHIM.
Given the limited resolution of our data, we also  
take a more conservative approach and provide upper limits according to \eqref{eq:OmegaWHIMXUL}, making use of the \eagle\ estimates of the expected column densities
of \ovii\ and \oviii. These upper limits remain valid even in the case that
some of the possible absorption lines tentatively identified in this paper were
spurious.

\subsubsection{Constraints assuming X--ray detections}
According to the discussion of Sec.~\ref{sec:cosmo-detections}, we consider the 7 
systems with a possible \ovii\ absorption line, and the 8 systems with a possible \oviii\
absorption line. These are the systems marked with a "Y" in Table~\ref{tab:detections}.
We use Eq.~\ref{eq:OmegaWHIMX} separately for the two ions, and obtain
the following results:
\begin{equation}
\ObXbFracEq = 
    \begin{cases}
      0.83^{+3.99}_{-0.62} \times\,\left(\dfrac{\fion}{1}\right)^{-1} \left(\dfrac{A}{\text{0.2~Solar}}\right)^{-1} \mathrm{(\ovii)} \\[10pt]
     0.79^{+3.08}_{-0.50} \times\,\left(\dfrac{\fion}{0.5}\right)^{-1} \left(\dfrac{A}{\text{0.2~Solar}}\right)^{-1} \mathrm{(\oviii)} 
    \end{cases}
    \label{eq:OmegaDetection}
\end{equation}
In Eq.~\ref{eq:OmegaDetection}, the column densities and distances for the \xmm\ and \chandra\ data are added together to give a combined result for each ion, with the
relevant quantities being displayed ion Table~\ref{tab:OMEGAWHIM}. In particular, distances to each
source are obtained from their redshift, as explained in Sec.~5.3 of \cite{spence2023} to which
we refer the reader for further details, and then added for
all the sources in a sample.
The results are parameterized as a function of the unknown metal abundance, which was assumed 
to be 0.2 Solar as a reference value. They are also a function of the ionization
fraction, for which we used a reference value of 100\% ionization for \ovii, and 50\% ionization for \oviii, corresponding to the approximate peak ionization fractions in collisional equilibrium \citep[e.g.][]{mazzotta1998}. 

In these estimates, and in the ones that follow, we make no assumption on the source of ionization, i.e., whether collisional or photoionization. Rather, we simply parameterize the results as a function of the ionization fraction. It
is expected that, if the absorption lines originate from low-density WHIM environments, photoionization plays
an important role. Compared to collisional ionization equilibrium, photoionization typically results in broader ionization peaks for \ovii\ and \oviii\ \citep[e.g.][]{kang2005}. This makes it such that there is a broader range
of temperatures where there is substantial \ovii\ and \ovii, and also a range ($\log T \mathrm{(K)}\simeq  5.5-6$) where there is also significant \ovi.

\begin{table}
    \centering
    \begin{tabular}{l|llll}
    \hline
    \hline
    Sample & $\log D$ & $N$ & $N_S^{U}$\\[5pt]
           & (Mpc) & ($\times 10^{17}$~cm$^{-2}$) & ($\times 10^{17}$~cm$^{-2}$)  \\[5pt]
           \hline
     \expandableinput Omegas-HandDetections-ALL.tex
    \hline
    \hline
    \end{tabular}
    \caption{Summary of { data} for the measurement of the cosmological density of X--ray WHIM baryons from the 7 \ovii\ and 8 \oviii\ possible detections marked with a "Y" in  Table~\ref{tab:detections}.}
    \label{tab:OMEGAWHIM}
\end{table}

\subsubsection{Constraints using upper limits}

If we ignore the possible detections and use instead only the upper limits to the
ion column densities of all systems in our sample, the cosmological results according to \eqref{eq:OmegaWHIMXUL} 
can be calculated from the data  presented in Table~\ref{tab:OmegaWHIMXUL}, where $D$ is the sum of all distances, $\Delta z$
the total redshift path sampled, and $\E[N_{ion}^+]$ the \eagle--based 
column density upper limits. Summing these upper limits over the \ovi\ and \hi\ priors,
and over the two separate instruments, we obtain the following overall upper limits, { at 90\% confidence,} to the
cosmological density of the two ions:

\begin{equation}
\ObXbFracEq \leq 
    \begin{cases}
       2.65 \times\,\left(\dfrac{\fion}{1}\right)^{-1} \left(\dfrac{A}{\text{0.2~Solar}}\right)^{-1} \mathrm{(\ovii)} \\[10pt]
      1.65 \times\,\left(\dfrac{\fion}{0.5}\right)^{-1} \left(\dfrac{A}{\text{0.2~Solar}}\right)^{-1} \mathrm{(\oviii)} 
    \end{cases}
    \label{eq:OmegaUL}
\end{equation}

For simplicity, we only report the upper limits based on the sensitivity method, since the
results for the Feldman--Cousins method are equivalent, as can be seen from Table~\ref{tab:OmegaWHIMXUL}. For these upper limits, as explained
in Sec.~\ref{sec:cosmo-method}, we make use of the \eagle\ expected column densities,
based on the measured upper limits. Results of \eqref{eq:OmegaUL} are independent from
those in \eqref{eq:OmegaDetection}, and therefore they remain valid even in the case that some
of the possible absorption--line detections were  inaccurate.

\begin{table}
    \centering
    \begin{tabular}{l|llll}
\hline
\hline
Sample & $\log D$ & $\Delta z$ &  \multicolumn{2}{c}{$\log \E[\nion^+]$} \\
        &         &            & Sens & FC \\
       & (Mpc) &  & \multicolumn{2}{c}{(cm$^{-2}$) }   \\[3pt]
\hline
 \expandableinput Omegas-ALL.tex
\hline
\hline
    \end{tabular}
    \caption{Expectations of column densities based on \eagle\ simulations and using upper limits to the observed column densities of the 
    of absorption lines, and { other} relevant survey parameters. }
    \label{tab:OmegaWHIMXUL}
\end{table}

\section{Cosmological interpretation}
\label{sec:interpretation}

\subsection{The missing baryons problem}
\label{sec:missingBaryons}

In a general--relativity Friedmann--Robertson--Walker cosmological model, the expansion of the Universe can
be described by three density parameters that obey the relationship $\Omega_m+\Omega_{\Lambda}+\Omega_{k}=1$,
which are respectively the contributions to the energy budget from matter (both dark matter and ordinary baryonic matter), a cosmological constant or more generally dark energy, and the curvature of the universe \citep[e.g.][]{carroll1992}. Of particular interest to this project is the contribution
of ordinary baryonic matter to the cosmological density of matter \Om, which is referred to as \Ob.

There are several independent pieces of observational evidence suggesting that the cosmological density
of baryons is approximately \Ob=$0.05$, or 5\% of the total cosmological density of
energy. First, models of big--bang nucleosynthesis \citep[e.g.][]{burles2001} agree with
observational measurements of deuterium--to--hydrogen ratios \citep[e.g.][]{kirkman2003},
indicating a value of $\Omega_b h^2 \simeq 0.02$, with $h\simeq 0.7$ the normalized value
of the Hubble constant \citep[e.g.][]{freedman2001,freedman2012, riess2019, riess2016,planck2020}.~\footnote{Although there is 
tension among different measurements of the Hubble constant, the differences are at the percent level, and do not have a significant effect on the missing baryons problem.}
Second, the analysis of cosmic microwave background (CMB) anisotropies by the \emph{Wilkinson Microwave
Anisotropy Probe} (WMAP) and by the \emph{Planck} missions concur on a similar value of \Ob, when used
in conjunction with other observations \citep[e.g.][]{planck2020,dunkley2009}.
Third, and more directly related to the type of analysis performed in this paper,
high--redshift observations of the hydrogen Lyman--$\alpha$ forest \citep[e.g.][]{rauch1998, weinberg1997} provide direct spectroscopic evidence that early--universe baryons
are in the amount predicted by big--bang nucleosynthesis models. More recently,
measurements of the dispersion of fast radio bursts were able to constrain the cosmological
density of baryons to a value of \Ob$\simeq0.05$, again consistent with big--bang nucleosynthesis and
CMB measurements \citep[e.g.][]{yang2022,macquart2020}.

Since the early seminal work of \cite{cen1999}, different generations of numerical simulations 
have consistently predicted that a significant portion 
of the present--day baryons are in the warm--hot intergalactic medium  at approximately $\log T(\mathrm K) = 5-7$ \citep[e.g.][]{branchini2009,schaye2015,mernier2017, tuominen2021}. 
The lower temperature range of the WHIM has been effectively and conclusively probed by
far--ultraviolet surveys of absorption lines
in the spectra of background sources, finding approximately \emph{one half} of the expected baryons \citep[e.g.][]{shull2012,tilton2012,danforth2016}. The growing consensus is that the remaining low--redshift baryons are in the
higher--temperature portion of the WHIM that is most readily observable in X--rays.
A review of the literature on X--ray absorption lines in search of the missing WHIM baryons 
was provided in \paperOne, where we have shown that at present there has been no conclusive 
determination of the cosmological density of X--ray absorbing WHIM baryons.

The analysis presented in this paper and in \paperOne\ was designed as 
a comprehensive census of the X--ray absorbing intergalactic medium at low redshift. 
The large
redshift path and number of sources make this a 
representative sample of the local universe, and the results
obtained in Sec.~\ref{sec:cosmo} are now used to address the missing baryons problem. We begin 
with a detailed account of the major sources of systematic error affecting our measurements.

\subsection{Sources of systematic uncertainty}
\label{sec:systematics}

This section discussed sources of systematic error in the estimate of the
cosmological density of baryons from the X--ray absorption line measurements. Sources of
systematic uncertainty associated with the measurements of the line themselves
were presented in Sec.~4.2 of \paperOne, and they are not discussed in this paper.

\subsubsection{Temperature and abundance}
\label{sec:systematics-TA}
The X--ray data we have analyzed do not provide meaningful constraints on the 
temperature and abundance of the absorbers for most of the sources. In the case of the 
brightest sources,
constraints on the temperature could be obtained by line ratios of ions of the same
elements, similar to what was done in \cite{ahoranta2020} or \cite{ahoranta2021}{, and could reveal whether multi-phase gas is required to explain the measured absorption}.
Given the limited resolution of the X--ray data for several of the sources in this sample,
we limit ourselves to parameterize the results as a function of an assumed temperature and abundance{, regardless of the origin of the ionization}. 

Specifically, we chose to parameterize our results in \eqref{eq:OmegaDetection} and \eqref{eq:OmegaUL} using the the largest expected values for the ionization fraction ($\fion$). This means that, if the temperature
of the absorber is significantly different from that at the peak of the  ionization curves, the
estimates would be systematically \emph{higher}. For example, as illustrated by the CIE curves of
\cite{mazzotta1998} and also shown in Fig.~2 of \cite{spence2023}, a WHIM system at 
$\log T(K)=6.3-6.5$
would be near the peak of the \oviii\ CIE ionization curve, and the nominal
value assumed for \oviii\  in this paper would be accurate. However, 
at the same temperatures the ionization fraction of \ovii\
would be in the range of $\sim$10-50\%, and therefore the column density of \ovii\ would be 2-10 times larger than assumed in our estimates. It is therefore clear that an order--of--magnitude systematic error  is possible, given 
our substantial ignorance of WHIM temperature from these X--ray data.
As an illustration of this effect, Table~\ref{tab:ionization}
reports values of \ObXb\ for selected values of the ionization fraction of \ovii\ and \oviii\ in CIE according to \cite{mazzotta1998}. The analysis shows that temperatures that are significantly away from peak ionization would result in extreme values of the inferred baryonic density. 
\begin{table}
    \centering
    \begin{tabular}{l|ll|ll}
    \hline
    \hline
    $\log T\,(\mathrm K)$ & \multicolumn{2}{c}{\ovii} &  \multicolumn{2}{c}{\oviii} \\
                & $\log f_{\text{ion}}$ & \ObXb & $\log f_{\text{ion}}$ & \ObXb\\
    \hline\\
    6.0 & -0.01 & $0.84\pm^{4.05}_{0.63}$ & -2.06 & $45.04\pm^{175.60}_{28.51}$ \\[5pt] 
6.1 & -0.03 & $0.88\pm^{4.25}_{0.66}$ & -1.25 & $6.99\pm^{27.26}_{4.43}$ \\[5pt] 
6.3 & -0.29 & $1.62\pm^{7.80}_{1.21}$ & -0.38 & $0.95\pm^{3.71}_{0.60}$ \\[5pt] 
6.5 & -1.13 & $11.27\pm^{54.20}_{8.42}$ & -0.50 & $1.24\pm^{4.84}_{0.79}$ \\[5pt] 
7.0 & -3.64 & $3623.08\pm^{17416.98}_{2706.40}$ & -1.81 & $25.33\pm^{98.75}_{16.03}$ \\[5pt] 

    \hline
    \hline
    \end{tabular}
    \caption{Ionization fractions in CIE \protect\citep{mazzotta1998} and associated values of the cosmological density of X--ray absorbing baryons (for 0.2 Solar abundances), for selected 
    temperatures.}
    \label{tab:ionization}
\end{table}

As already discussed in Sec.~\ref{sec:results}, there is no guarantee that the absorbing plasma is in collisional equilibrium. In fact, the ionization balance is expected to be
affected by photoionization  in the more diffuse WHIM environments \citep[e.g.][]{nelson2018}. If photoionization
is significant, the ionization fraction is no longer simply a function of temperature, but it becomes
also a function of the plasma density and of the intensity of the ionizing field. The substantial ignorance of the ionization fraction is what drove our choice to parameterize the results with the most
conservative value of the ionization fraction.

We also parameterized our results assuming a 0.2~Solar abundance of oxygen, with
the oxygen--to--hydrogen abundances of \cite{anders1989}, consistent
with expectations from current cosmological simulations \citep[e.g.][]{martizzi2019, nelson2018,rasia2008} and observations
of the outskirts of galaxy clusters \citep[e.g.][]{degrandi2001, mernier2017}, where WHIM filaments are expected to converge. The main source of systematic error
associated with this assumption is that the WHIM abundance of oxygen can be different from the
nominal value of 0.2~Solar by more than a factor of 2, therefore making the estimates
uncertain by a similar factor. An oxygen abundance that is close to Solar would provide a
significant reduction in our estimates, although such high abundances are not typically seen in
simulations of WHIM filaments \citep[e.g., see Fig.~9 of][]{martizzi2019}.

\begin{figure*}
    \centering
    \includegraphics[width=6.5in]{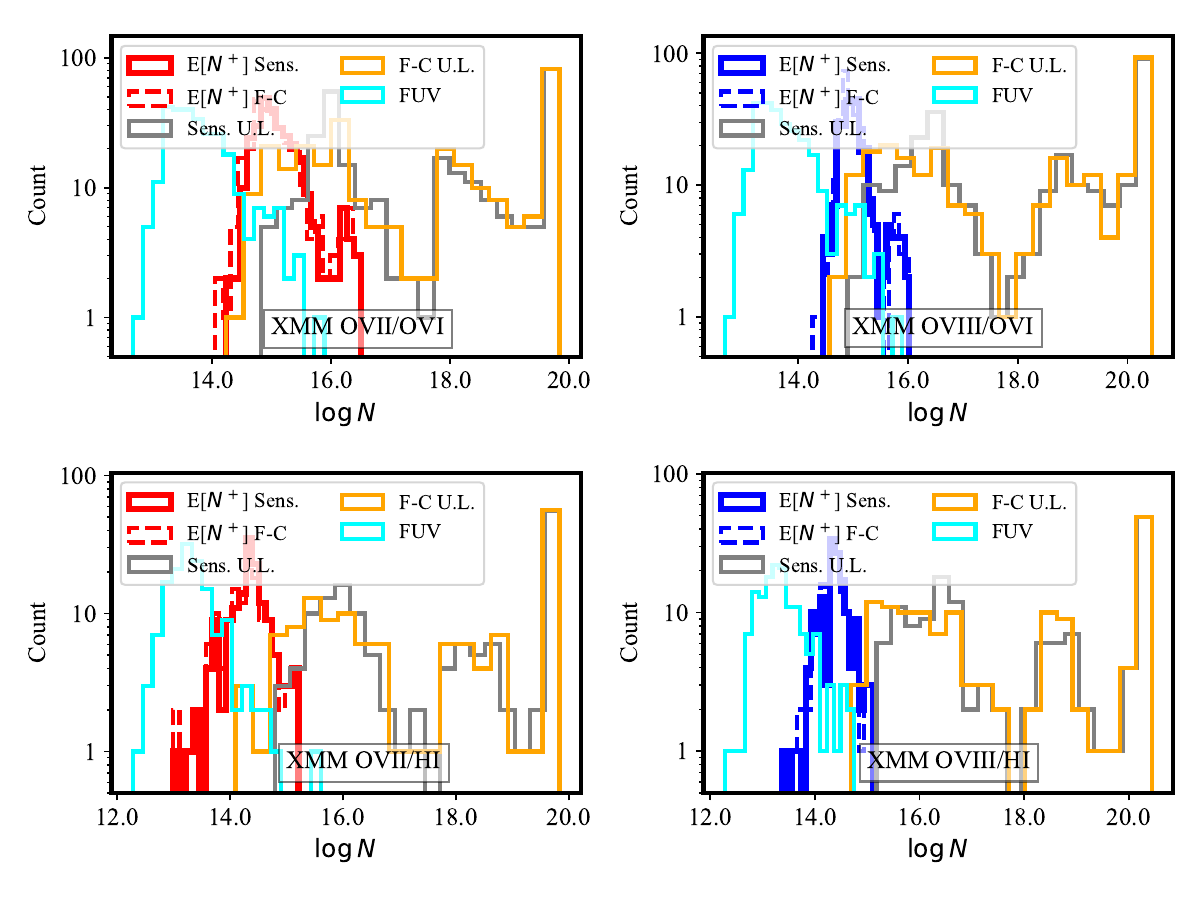}
    \caption{Distributions of: FUV column densities (cyan) from \protect\cite{tilton2012}
    and \protect\cite{danforth2016}; expected column densities        $\E[N^+]$ for all absorption line systems
    that are summed in \eqref{eq:nionMinus}, calculated using the "sensitivity" and the Feldman--Cousins methods (color--coded); and  X--ray upper limits  to the measurable column densities for the two methods (orange and grey), 
    for the \xmm\ data. }
    \label{fig:logN-XMM}
\end{figure*}

\begin{figure*}
    \centering
    \includegraphics[width=6.5in]{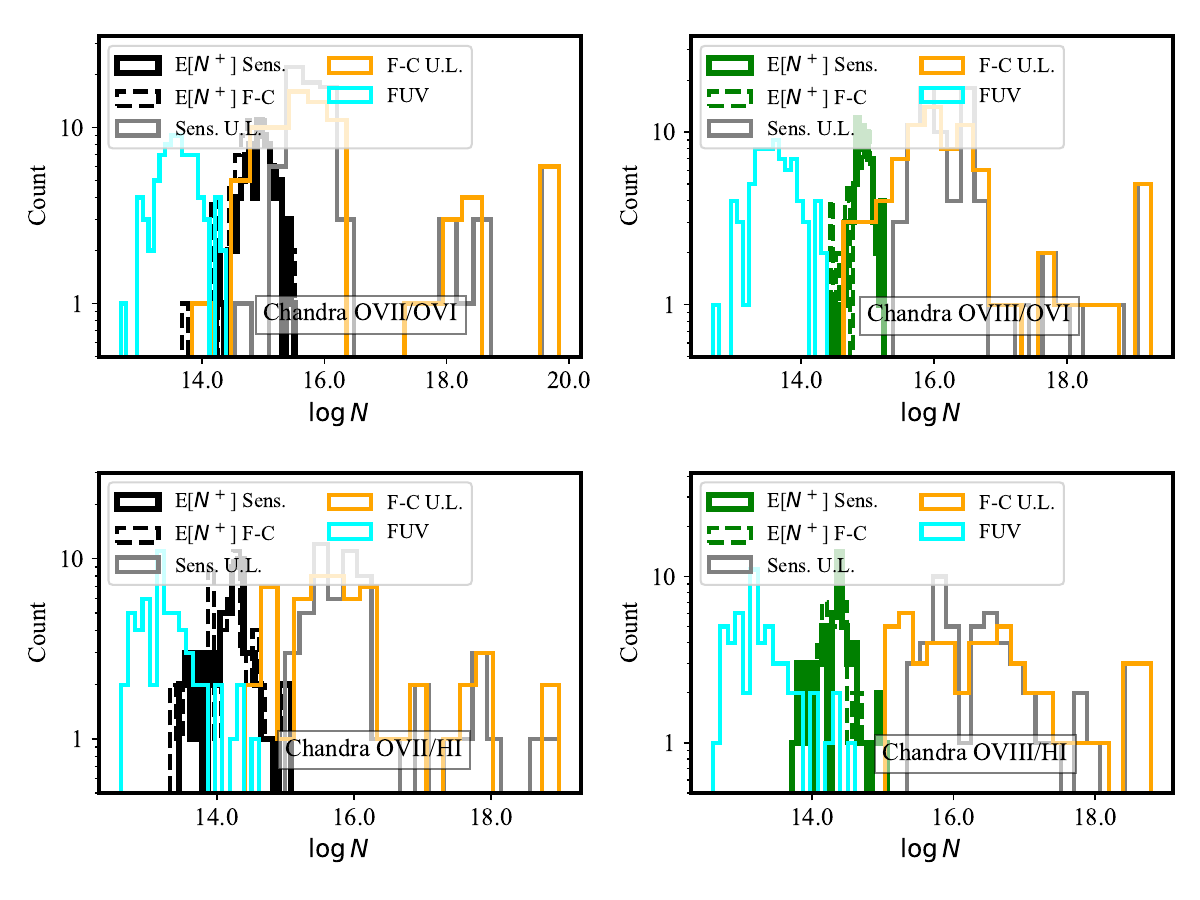}
    \caption{Same as Fig.~\ref{fig:logN-XMM}, for the \chandra\ data.}
    \label{fig:logN-Chandra}
\end{figure*}

\subsubsection{The $b$ parameter}
\label{sec:systematics-b}
Similar to the case of temperature and abundance, our analysis is blind to the
Doppler $b$ parameter of the absorption line. As discussed in Sec.~\ref{sec:lineProfile}, this
parameter is crucial for an accurate estimate of the column density from the measured
equivalent width of the lines, as illustrated also in Fig.~\ref{fig:COG}. A lower
limit to the $b$ parameter could be set according to \eqref{eq:b}, if the temperature of the
absorber was known. Larger $b$ values can occur in the presence
of non--thermal broadening that can be caused by velocity structure or turbulence 
in the WHIM \citep[e.g.][]{richter2006,schmidt2021}. Substantial non--thermal
broadening is in fact present in some of the FUV spectra used as priors
for this search \citep[e.g][]{danforth2016}. 

Given that our possible detections in Table~\ref{tab:detections} typically correspond to
equivalent widths $\log \Wl(\text{mA}) \gg 0$, and often $\log \Wl(\text{mA}) \gg 1$, Fig.~\ref{fig:COG} shows that
the impact of the assumed $b$ parameter is substantial. We repeat the analysis leading to
the estimates of \eqref{eq:OmegaDetection} for a lower value of $b=30$~\kms, corresponding to predominantly thermal broadening only,
and a larger value of $b=200$~\kms, corresponding to a significant contribution from non--thermal
broadening, and obtain the following results:
\begin{equation}
\ObXbFracEq= 
    \begin{cases}
    9.9\pm^{10.25}_{5.7} (b=30)\,,\, 0.165\pm^{0.159}_{0.060} (b=200)\, \mathrm \ovii  \\[5pt]
    29.9\pm^{37.0}_{22.5}  (b=30)\,,\, 0.435\pm^{0.252}_{0.186} (b=200)\, \mathrm \oviii,
    \end{cases}
\end{equation}
{ subject to the same scaling with abundances and ionization fraction as in Eq.~\ref{eq:OmegaDetection}.}
It is therefore clear that the $b$ parameter has a major effect on the determination
of the cosmological density of WHIM baryons. In particular, a substantial contribution
from non--thermal broadening would significantly reduce the estimates. On the other hand,
a small $b$ parameter would significantly increase the estimates, as the departure from
the linear \cog\ regime would occur for lower values of the equivalent width.

On the other hand, the upper-limits estimates in \eqref{eq:OmegaUL} are only minimally affected
by the estimate of the $b$ parameter, since they are based on the \eagle\ estimates.
In fact, the measured upper limits --- which indeed change significantly as a function of $b$ ---
only enter the calculation as a limit of integration, and the resulting upper limits
for $b=30$ or 1000~\kms\ are changed only by a few percent relative to those for $b=100$~\kms.

\subsubsection{Estimates from \eagle}
This source of systematic error only affects the upper limits, since no use of the
\eagle\ simulations was made in the estimate of \eqref{eq:OmegaDetection} based on the positive detections.
The method used in this paper to calculate \ObX\ is based on the use of \eagle\ simulations
to predict the average amount of \ovii\ and \oviii\ along a given sightline with a prior FUV detection (see Sec.~\ref{sec:cosmo}, and Sec.~5 of \citealt{spence2023}).
The key assumptions of this method are that (a) the FUV data have identified all possible
redshifts where associated X--ray absorption can be found; and (b) the $\log N$--$\log N$
scaling relation between the FUV and X--ray column densities are accurate in predicting \ovii\ and \oviii\ column densities associated with the \ovi\ and \hi\ FUV absorbers (these
releationships were identified by \citealt{wijers2019} and were studied in \citealt{spence2023}).
Both of these assumptions are reasonable, but basically untested, given the paucity of X--ray absorption line systems and their limited resolution.

For assumption (a), we point out that X--ray absorbers need not necessarily be associated with
FUV systems. For example, systems with $\log T(K) \gtrsim 6$ would have minimal or no \ovi\ absorption, and therefore they would be missed by a search based on \ovi\ priors, like the present one. Therefore it is possible that our search has missed hotter absorption--line systems at other redshifts.
Regarding assumption (b), the use of \eagle\ scaling relations to estimate upper limits to the column density of \ovii\ and \oviii\ is also a convenient tool, yet one that is largely untested. In particular, the distributions of Figs.~\ref{fig:logN-XMM}
and \ref{fig:logN-Chandra} show that sightlines with \hi\ priors predict substantially
\emph{lower} X--ray column densities than those with \ovi\ priors, according
to the \cite{wijers2019} analysis of the \eagle\ simulations. 

Other methods could be used to set an upper limit to the column density along
a specific sightline. A direct method would be that of using the measured upper limit \NUL\
itself, as already discussed in Sec.~\ref{sec:cosmo-method}, which would correspond to ignoring the information provided by the column density of the FUV prior, and the predictions based on \eagle. That method would give substantially larger upper limits to \ObX, given that the \eagle\ expectations for the unobserved column densities
are substantially lower than the raw X--ray upper limits, see Figs.~\ref{fig:logN-XMM}
and \ref{fig:logN-Chandra}.
Moreover, as already remarked in \cite{spence2023}, our method of using FUV priors implies
that we consider \emph{all} absorption line systems and sum the expected column densities
according to \eqref{eq:nionMinus}, as per assumption (a). 
In the absence of FUV priors, one could set the amount of the unobserved column density to
the value of the most stringent upper limit along that sightline. However, that method
would ignore both the observed FUV priors, and the predictions based on \eagle\ simulations.

\subsubsection{Double counting of absorption systems}
\label{sec:systematicsDoubleCounting}
There are two issues concerning the possible double- or multiple--counting of absorption--line system. The first concerns the counting of FUV systems used as priors, and the second
the counting of the X--ray systems themselves.

Certain sightlines have a large number of \ovi\ and \hi\ FUV absorption line systems, which
we treat as independent for the sake of predicting the associated X--ray column densities, i.e., we make X--ray predictions for each and then we sum them.
It is possible that certain FUV systems are in fact associated with the same absorber,
and therefore it would be appropriate to first add the prior FUV column densities
of all systems that are physically associated, and then predict just one X--ray column density per system, using the summed FUV prior column density. The FUV data used in this paper do not provide an unequivocal way to determine such
association, and therefore we did not attempt it. However, we find that the predicted
$\E[\nionj^+]$ has a positive { correlation with}
 the column density of the FUV prior
(see Fig.~8 of \citealt{spence2023}), so that the order
in which to add column densities (adding the FUV priors first, or adding the 
expected X--ray column densities for each FUV system) is not a major source of
systematic uncertainty.

More important is the possibility of double--counting the detection of possible
X--ray absorption lines. For example, among the systems with possible detections, systems 3 and 4 for \es\ are obtained from FUV priors at $z=0.18759$ and $z=0.18775$ (from \citealt{danforth2016} reported
in Table~2 of \paperOne). These two redshifts are indistinguishable at the resolution of
either \xmmshort\ or \chandra, and therefore it would seem appropriate to consider just one of the
two possible detections towards the determination of \ObX. Since the two lines have similar estimated column densities, ignoring the first leads to an estimate of \ObX\ for \ovii\ that is reduced by 0.8\%, i.e., a negligible effect. In fact, that estimate is dominated by the larger column densities of other systems.

Another case is that of the  \xmm\ 3C273 absorber 21 for \ovii\ at $z=0.0902$, which has a corresponding possible detection of \oviii\ at the same redshift (system 77, \chandra\ data). In this case, it is also possible, if not likely, that
both absorption features are part of the same system. This is the only case in our sample of a
system with more than one possible detection, as also discussed in Sec. 3.4 of \paperOne\  with
a caveat on a possible Galactic origin discussed in Sec.~\ref{sec:confusionGalactic}, and therefore raises the issue of possible double--counting of baryons between the \ovii\ and \oviii\ estimates. 
The spectra for the \ovii\ \xmm\ and the \oviii\ \chandra\ absorption features
were shown respectively in Figures~5 and 6 of \paperOne, to which the reader is referred for additional
details on this source, and where the overall significance of detection of both lines with both
instruments was also discussed.~\footnote{An earlier and more in-depth analysis of the 3C273 source using the same data as 
in this paper was presented in \cite{ahoranta2020}, where the challenges in disentangling the possible
\ovii\ at $z=0.0902$ from Galactic \oi\ were discussed.}
This problem is akin to that
discussed by \cite{danforth2016} regarding the estimates of cosmological baryon densities 
calculated from individual FUV ions (e.g., \ovi\ or \nv). Given the limited size of 
the detections in our analysis, and the unavailability of temperature information, we do not attempt to combine the \ovii\ and \oviii\ estimates of the cosmological density of baryons into
one single estimate for the X--ray absorbers { (same as \citealt{danforth2016})}. We point out that, if the 3C273 $z=0.0902$
system has a temperature that is somewhat intermediate to the peaks of the \ovii\ and \oviii\
ionization curves, the corresponding total baryon density implied might in fact be even larger than
that assumed by the two individual estimates, because of the lower implied ion fraction, see \eqref{eq:Omega}. Therefore
we simply  caution that the possible double--counting of \ovii\ and \oviii\
 absorbers is such that the overall estimate of \ObX\ may be different from the
 sum of the two contributions by \ovii\ and \oviii, and a more in depth analysis is
 required to combine the estimates. Such analysis{, which was done for Ton~S180 in \cite{ahoranta2021},} is not performed in this paper.

\subsubsection{Determination of detection status}
\label{sec:sysDetection}
Perhaps the most fundamental source of systematic error in the estimates of $\OWHIMX$ is the determination of
which systems have a positive detection of an absorption line. The criteria for detection
are explained in Sec.~\ref{sec:cosmo-detections}, where we reduced the 33+1 absorption--line features with $\Delta C \geq 6.6$ to just 7 \oii\ and 8 \oviii\ systems, based on a number
of considerations that include the possibility of Galactic or intrinsic absorption. Some
of these 15 systems were also marked as being of an uncertain nature, i.e., the 3C273 \ovii\ system 23 (possible intrinsic absorption), and \es\ \oviii\ systems 8 and 9 (absorption lines are in a region of reduced efficiency).

To illustrate the effect of this systematic uncertainty, we repeated the same analysis as
in \eqref{eq:OmegaDetection} with the exclusion of these systems, and obtain the following results:
\begin{equation}
\ObXbFracEq= 
    \begin{cases}
    0.82\pm^{3.98}_{0.62}\, { \times\,\left(\dfrac{\fion}{1}\right)^{-1} \left(\dfrac{A}{\text{0.2~Solar}}\right)^{-1}} (\mathrm \ovii)  \\[5pt]
    0.68\pm^{3.04}_{0.46}\, { \times\,\left(\dfrac{\fion}{0.5}\right)^{-1} \left(\dfrac{A}{\text{0.2~Solar}}\right)^{-1}} (\mathrm \oviii).
    \end{cases}
\end{equation}
As expected, the results are virtually identical to those of \eqref{eq:OmegaDetection}, since the
estimates are dominated by systems with larger estimated column density. This source of
systematic error is therefore closely associated with the ones described in Sec.\ref{sec:systematics-TA} and \ref{sec:systematics-b}, which dominate the systematic error
budget.

{ An alternative means to address this systematic is to raise the
significance of detection, in order to reduce the possibility of using false positives for the cosmological estimates. If we consider only systems that have $\Delta C \geq 9$, which corresponds to a $p$--value of 0.0027
and is equivalent to a ``3-$\sigma$ detection'', namely systems 308 (Ton~S180), 23 and 21 (3C~273) for \ovii/\ovi, and systems 114 (Mkn~421), 8 and 9 (1ES~1553) for \oviii/\ovii, all from \xmm\ observations, we obtain:
\begin{equation}
\ObXbFracEq= 
    \begin{cases}
    0.56\pm^{1.83}_{0.41}\,  { \times\,\left(\dfrac{\fion}{1}\right)^{-1} \left(\dfrac{A}{\text{0.2~Solar}}\right)^{-1}} (\mathrm \ovii)  \\[5pt]
    0.11\pm^{0.05}_{0.04}\, { \times\,\left(\dfrac{\fion}{0.5}\right)^{-1} \left(\dfrac{A}{\text{0.2~Solar}}\right)^{-1}}(\mathrm \oviii).
    \end{cases}
\end{equation}
In particular, the substantial reduction in the the nominal \oviii\ estimate is due to the exclusion of the 1ES~1028 and 3C~273 \oviii\ systems that feature the largest estimated column density (see Table~\ref{tab:detections}). It is worth noting that
the \oviii\ 3C~273 system at $z=0.0902$ has the same redshift as the 3C~273 \ovii\ system used in the \ovii\ estimate. As discussed above in Sec.~\ref{sec:systematicsDoubleCounting}, this redshift system for 3C~273 is the only one in the sample to have both \ovii\ and \oviii\ detections at the 99\% level, and therefore both absorption lines are in fact likely to be real. 
Nonetheless, even with this more stringent criterion, the estimated baryon budget from the X--ray absorbing WHIM remains well consistent with
the amount required for the resolution of the missing baryons problem. \
}

\subsubsection{WHIM vs. WCGM origin of the absorption}

In this paper we refer to WHIM as any gas in the temperature range that is suitable to 
yield X--ray absorption lines, regardless of its density. Numerical simulations
show that 
intergalactic gas in the $\log T(K)=5-7$ range can be conveniently divided into
low--density, high--entropy gas that resides predominantly in filaments, and often referred to as WHIM proper; and high--density, low--entropy gas that preferentially resides in the circum--galactic
medium, referred to as warm circum--galactic medium \citep[WCGM, e.g.][]{galarraga2020,gouin2022, gouin2023}. 
Since both the FUV data used for redshift priors, and the X--ray data, are insensitive to
the density of the absorber, we do not address this distinction in this paper. Our effort to
remove possible absorption line systems that are at a small impact parameter from known
massive galaxies suggests that the systems with positive detection are likely
to be associated with the truly diffuse filamentary WHIM. Further evidence in favor of a diffuse WHIM origin
is provided in the following section.

\subsection{Correlation with galaxy filaments}

Given that most of the WHIM is expected to reside in cosmic filaments, we performed an analysis of
the correlation between the possible absorption systems from Table~\ref{tab:detections} and known filament catalogs. The main observable tracer of the filamentary network are galaxies, with different methods choosing different approaches to extract the long, elongated filaments from discrete galaxy distributions. Thus, while different filament catalogs may have similar statistical properties (e.g. number of galaxies they contain), they will not trace
the exact same filaments in space. Since the goal of our analysis is 
to find the closest filaments in space to the detected absorbers, we searched different filament catalogs in order to minimize the uncertainties related to a single catalog. 

\subsubsection{Choice of filament catalogs}

For a filament catalog to be as reliable and complete as possible, it must be constructed using a three-dimensional sample of spectroscopic galaxies with very small uncertainties in their redshift distances. Currently, the most complete filament catalogs have used data releases from the Sloan Digital Sky Survey \citep[SDSS,][]{york2000}, a large spectroscopic survey containing a vast number of galaxies. Among the tested and complete filament 
catalogs we chose three, based on their public availability and readiness to be used for this analysis. 
First, the catalog by \citet{tempel2014a} was constructed using the Bisous method \citep{stoica2007,tempel2014a,tempel2016}, and it contains 15,421 filaments. Second, \citet{malavasi2020} used the DisPerSE method \citep{sousbie2011a,sousbie2011b} to construct a set of catalogs, with different choices of parameters for their analysis. Among those DisPerSE catalogs we used that
based on the LOWZ+CMASS SDSS data, with no prior smoothing and $3\sigma$ persistence cut. This selection provides the largest number of filaments, 63,390 in total, allowing for smaller structures (i.e., `tendrils') to also be present. The third catalog we chose was by \citet{carronduque2022} with the  subspace-constrained mean shift \citep[SCMS,][]{chen2015} filament finder. Given our redshift coverage, we used Block 1 of this catalog which covers redshifts between $0.05<z<0.045$. 
These three catalogs therefore use similar SDSS data, but substantially different methods for the identification of
filaments.

\subsubsection{Results of the search}
Our selection of filament catalogs is restricted to the area covered by the SDSS. From our possible \nDet+1 detections
listed in Table~\ref{tab:detections}, a total of six sources and 13 systems are within the area covered by the SDSS data, listed in Tab.~\ref{tab:fil_distance}. For each system we measured the impact parameter to the closest Bisous, DisPerSE and SCMS filament, within $\Delta z \approx 0.0034$ of the absorbers redshift. These absorption systems reach a redshift of $z \approx 0.38$, which is beyond the redshift range covered by the Bisous filaments ($z < 0.15$), so only a few of the systems
can be searched for this type of filaments. At the same time, due to the very low redshift ($z\approx0.01$) of system 114 (Mrk 501), the SCMS and DisPerSE methods do not provide reliable filaments given the low number density of galaxies. Therefore, we obtained distances to Bisous filaments for 5 systems, and distances to DisPerSE and SCMS filaments for 12 systems. Table \ref{tab:fil_distance} shows the distance to the closest filament for each of the 13 absorption-line systems
in the SDSS volume.

\begin{table}
    \centering
    \begin{tabular}{c|cc|cc}
    \hline
    \hline
    Source & \multicolumn{2}{c}{Absorption System} &  \multicolumn{2}{c}{WHIM Filament}\\ 
         & ID & $z$ & Distance (Mpc) & Type\\
    \hline
    3c273 & 21 & 0.0902 & 1.69 & DP \\
    3c273 & 23 & 0.1466 & 16.53 & DP \\
    1es1553 & 4 & 0.1878 & 8.24 & DP \\
    1es1553 & 5 & 0.1876 & 8.24 & DP \\
    1es1553 & 6 & 0.1898 & 8.24 & DP \\
    1es1553 & 8 & 0.3113 & 38.69 & SCMS \\
    1es1553 & 9 & 0.3787 & 4.29 & DP \\
    pg1211 & 180 & 0.0512 & 0.59 & SCMS \\
    1es1028 & 3 & 0.3373 & 15.02 & DP \& SCMS \\
    mrk421 & 114 & 0.0101 & 2.2 & Bis \\
    3c273 & 77 & 0.0902 & 1.69 & DP \\
    pg1116 & 75 & 0.0838 & 1.5 & Bis \& SCMS \\
    pg1116 & 38 & 0.1337 & 3.13 & DP \\
    \hline
    \hline
    \end{tabular}
    \caption{Distance to closest filament for absorption-line systems in Table~\ref{tab:detections} with SDSS coverage. DP and Bis correspond to DisPerSE and Bisous filaments, respectively.}
    \label{tab:fil_distance}
\end{table}

Simulations predict a rather steep decrease in baryon density as a function of distance from filament spines \citep[e.g.][]{galarraga2021,galarraga2022,tuominen2021}, with baryon densities and temperatures reaching cosmic mean values at distances of a few to 10 Mpc. Moreover, \citet{tuominen2023} found that $\approx 68\%$ of the intergalactic oxygen lies within $\approx2$ Mpc of Bisous filament spines in the EAGLE simulation. 
It is also necessary to consider various sources of uncertainty in the location and density of possible WHIM filaments,
when considering whether the absorption lines can be generated by the WHIM in filaments. For example, 
feedback processes in simulations may have underestimated the reach of metals away from the filaments,
so that filaments may in fact reach further than indicated by the current generation of simulations \citep[e.g.][]{tuominen2023, galarraga2022}. In addition, the location of the detected filament spines using
any of these  methods are not exact, but rather the most likely path a true filament may be tracing \citep[e.g.][]{bonnaire2020}. In other words, there is an intrinsic uncertainty of $\sim 1$ Mpc or so in the actual location of the spine. 
In consideration of these points, we estimate that any absorber closer than $\sim 5$ Mpc to a filament spine could indeed be tracing filamentary WHIM, and with more certainty if the sightline is $\leq 2-3$~Mpc from the
filament spine. From our sample of absorbers in Table~\ref{tab:fil_distance}, we find that 7 of the 13 systems 
are closer than 5 Mpc to the spine of a known filament. 

\subsubsection{Correlation between filaments and absorption line systems}

Of the \nDet+1 systems listed in Table~\ref{tab:detections}, only 13 are within
the SDSS volume. 
Of the 7 systems with the closest impact parameter ($\leq 5$~Mpc) from a filament's spine,
systems 21/77 (3C~273), 114 (Mkn~421), 38 (PG~1116+215) and 8 (1ES~1553) were identified in
Table~\ref{tab:detections} as likely detection, with  system 75 (PG1116+215) and system 180 (PG~1211) having an
uncertain detection status. 
The remaining 6 systems with significantly larger distance from a detected filament (system 23 for 3C273, systems 4,5 and 6 for 1ES~1553 at the same redshift, system 8 also for 1ES~1553 at higher redshift, and system 3 for 1ES~1028) 
all have "likely" detections yet
there are no known filaments that can reasonably yield the measured
column density.
Finally, the remaining systems with reliable detection in Table~\ref{tab:detections} but without SDSS coverage
are system 308 (Ton~S180), system 267 (PKS~2155), systems 16 and 18  (both for H~1821); for them, no search 
for neighboring filaments was possible.

Overall, the fact that several of the likely X--ray absorption line systems lie within a few Mpc of a known filament
 is  evidence that \emph{at least a fraction} of the detected X--ray absorption lines {are} associated with
known galaxy filaments. The lack of known filaments towards the remaining systems may be interpreted either
with the need to improve the filament detection algorithms and/or the underlying galaxy data, or that certain
X--ray absorption lines are not truly associated with WHIM filaments. Given the limited statistics of our
X--ray systems with SDSS data, additional analysis is required to further test this association; such analysis
exceeds the scopes of this paper.

\subsection{{ Confirmation of the} resolution of the missing baryons problem}

The results presented in Sec.~\ref{sec:results} and the analysis of several
source of systematic errors in Sec.~\ref{sec:systematics} provide observational evidence that
\ovii\ and \oviii\ X--ray absorption lines in the spectra of distant quasars { indeed} trace
a cosmological density of baryons that is sufficient to solve the missing baryons problem{, consistent with
other independent probes (see Sec.~\ref{sec:missingBaryons})}.

The X--ray measurements presented in this paper do have several sources
of systematic uncertainty,  
due primarily to the lack of knowledge of fundamental physical parameters associated with
the absorbers, namely their temperature, abundance and density. As a result,
we have parameterized the $\OWHIMX$\ results presented in Eq.~\ref{eq:OmegaWHIMX} with
an ionization fraction and abundance that leads to a conservative estimate. Even so,
the estimated cosmological density of X--ray baryons is consistent with the value of $\OWHIMX \simeq \nicefrac{1}{2}$,
i.e., it is sufficiently large to solve the missing baryons problem.
Substantially larger
values for the cosmological density parameter would be obtained if the 
absorbers had a temperature that is different from that at the peak of the ionization curve, or if the metal abundances are less than the nominal value of 0.2~Solar.
The estimates presented in this paper also depend on the
velocity dispersion of the absorbing gas, which likewise we cannot measure. Even the use of a substantially larger value of the $b$ parameter compared to the
nominal  value of 100~\kms, as discussed
in Sec.~\ref{sec:systematics-b}, leads to smaller estimates of the baryon density
that nonetheless remain sufficiently large to bridge the missing baryons gap. 
Other sources of systematic
error analyzed in Sec.~\ref{sec:systematics} do not appear to be able to significantly alter this conclusion.

A complementary view of the missing baryons problem is provided by the upper limits
of Eq.~\ref{eq:OmegaWHIMXUL}. As remarked in Sec.~\ref{sec:systematics-b}, the upper limits are
virtually insensitive to the $b$ parameter, since they are based on the predicted
\ovii\ and \oviii\ column densities for the relevant FUV priors \citep{wijers2019}. Similar to the
considerations above for the detections, these limits are well consistent with
the cosmological density of baryons required to solve the missing baryons problem,
for virtually any range of reasonable temperatures and abundances of the absorbers.

\section{Conclusions}
\label{sec:discussion}

This paper has presented a comprehensive budget of the cosmological density of baryons
associated with the X--ray measurements of \ovii\ and \oviii\ absorption lines
in the spectra of \nSources\ \xmm\ and \chandra\ background quasars.
Of the \nsystems\ systems with 
FUV priors from \ovi\ and \hi\ BLA absorption line detections by \cite{tilton2012} and \cite{danforth2016}, we have identified a total of 33+1 absorption lines that satisfy the
formal 99\% criterion of detection for the \texttt{line} component we have used to model
the presence of absorption lines. Of these, we 
tentatively identified 7 \ovii\ and 8 \oviii\ likely absorption line systems associated with the warm--hot intergalactic medium.

The total redshift path covered by this search is $\Delta z \simeq 10$, larger than any of
the previous X--ray studies. The FUV survey by \cite{danforth2016}, used to identify FUV priors
for this work, covered a redshift path of $\Delta z=21.7$ towards 82 extragalactic sightlines,
and resulted in over 5,000 absorption--line detections of various ions, including the \ovi\ and \hi\ ions used as priors for this work. The  comparison between the number of absorption line detections
in the  FUV and in X--rays underscores the challenges associated with the current generation of
X--ray grating spectrometers, which are only sensitive to the largest values of the
possible WHIM column densities intercepted by the background radiation. For this reason,
to date most of the searches for X--ray absorption by the WHIM focused on the brightest
background sources \citep[e.g., recently by ][]{nicastro2018, kovacs2019, ahoranta2020, ahoranta2021, gatuzz2023}. In this project, on the other hand, we have opted to follow all
available X--ray sources in the background of the FUV priors provided by \cite{danforth2016} and \cite{tilton2012}.

Many of the previously reported WHIM absorption--line detections
have often been controversial because of the limited quality of the data, even
for the brightest sources with the longest exposure time, such as \es\ \citep[e.g.][]{spence2023, nicastro2018}. 
The results of this work highlight the same challenges in the detection of X--ray WHIM absorption lines 
as in previous searches 
\citep[e.g.][see \paperOne\ for a more complete list of references]{gatuzz2023,kovacs2019}. In fact,  as discussed
in Sec.~\ref{sec:selection}, the strongest absorption line detected in this work
has a $p$-value that is only  equivalent to a "4.2--$\sigma$ detection". It is therefore clear that,
in order to make further progress in the detection and characterization of X--ray absorption lines from the WHIM, instruments with substantially large effective area and resolving power at soft X--ray energies are needed.
Future mission concepts such as \emph{Lynx} \citep[e.g.][]{gaskin2019}, \emph{HUBS}  \citep[e.g.][]{zhao2024}, \emph{LEM} \citep{kraft2024} or \emph{Super--DIOS} \citep[e.g.][]{sato2022} have the potential
to provide order--of--magnitude improvements to the study of X--ray absorption from the WHIM.

Given the large redshift path and the large number of sources,
this study was designed to be representative of the amount of X--ray absorbing WHIM
in the low--redshift Universe.
The main limitations towards an accurate census of the X--ray WHIM is associated with the
lack of temperature and abundance information from the spectral fits to the X--ray data,
and in uncertainties in the velocity dispersion $b$  parameter.
The primary motivation for the use of a simple \texttt{line} model for the absorption line, compared to more physically motivated models used in other searches \citep[e.g.][]{spence2023, gatuzz2023, ahoranta2021}, was the overall feasibility and statistical accuracy of such a large project, as explained
in \paperOne. While a temperature analysis might have been attempted in a few cases,
as we did for Ton~S180 and 3C27s \citep{ahoranta2020, ahoranta2021}, it was not feasible for the whole sample, and was therefore not attempted. 

Despite these limitations, we have shown that the cosmological density of baryons
implied by the detection of \ovii\ and \oviii\ absorption lines are 
consistent with values $\OWHIMX \geq \nicefrac{1}{2}$, which is the value that
is required to solve the missing baryons problem{, consistent with other recent observations \citep[e.g.][]{macquart2020, yang2022}}. 
First, we used nominal values for the WHIM temperature that correspond to the peak of the
ionization curve in collisional ionization equilibrium, and a value of 0.2~Solar abundances that is towards the high end of the
expected range, so that the estimates of $\OWHIMX$ in \eqref{eq:OmegaDetection} are conservative.
Moreover, the analysis of systematics presented in Sec.~\ref{sec:systematics} show that even the use of an extremely large value for the
$b$ parameter --- which reduces the column density estimates according to the \cog\ analysis ---
still leads to  $\OWHIMX \simeq \nicefrac{1}{2}$. 
We therefore conclude that these data provide convincing evidence 
that 
X--ray absorption lines from  \ovii\ and \oviii\ { confirm the resolution of
the missing baryons problem},
and that the missing baryons indeed are in the $\log T(K) \simeq 6-7$ phase of the WHIM,
as predicted by simulations.

Moreover, we have provided evidence in support of the scenario that the X--ray absorption lines 
identified in this analysis
originate from
the WHIM in galaxy filaments. In fact, some of the absorption line systems with reliable detection
do have a nearby SDSS galaxy filament, at an impact parameter that may be consistent with the
inferred column density of oxygen \citep[e.g.][]{tuominen2023, galarraga2022}. It is
reasonable to ask whether certain X-ray absorption lines may originate from a different environment, perhaps
the WCGM of halos that are not directly associated with galaxy filaments. We find that none of the
absorption line systems that do not lie close to a known filament have a sufficiently nearby galaxy
that can readily explain the absorption column \citep[][\paperOne]{spence2024}. Therefore we do not find 
immediate direct evidence
for the association of X--ray absorption lines with the halo of individual galaxies. The eventuality that a few of the detections reported in this paper 
are spurious was treated as a source
of possible systematic error, and therefore it does not affect
our conclusions on the resolution of the missing baryons problem.

This analysis did not consider  certain previously reported detections towards some of the
brightest sources, such as \es\ \citep{nicastro2018, gatuzz2023, spence2023}, H~2356-309 \citep[e.g.][]{buote2009, fang2010, zappacosta2010, kovacs2019}, H~1821+643 \citep[e.g.][]{gatuzz2023}, Mrk~421 \citep{nicastro2005,rasmussen2007,yao2012}, PG1116+215 \citep{bonamente2016, bonamente2019b}, PKS~2155-304 \citep[e.g.][]{fang2002, cagnoni2004,fang2005, yao2009}, and more (see \paperOne\ for a more complete discussion). Several of those claimed detections, in fact, did not use an FUV prior, and no attempt was made to identify lines without an FUV prior
in this project. Serendipitous line searches \`{a} l\`{a} \cite{gatuzz2023} or \cite{nicastro2018} are
an alternative and complementary method to search for X--ray absorption from the WHIM, and 
they have shown the
potential to resolve the missing baryons problem \citep[e.g.][]{nicastro2018}. The identification of
additional serendipitous absorption--line systems adds further confidence to the conclusion
that the missing baryons are indeed in the hotter portion of the WHIM that gives rise to X--ray absorption lines 
in the spectra of background sources.

\section*{Acknowledgments}
DS and MB acknowledge support from NASA 2ADAP2018 program `\emph{Closing the Gap on the Missing Baryons at Low Redshift with multi--wavelength observations of the Warm--Hot Intergalactic Medium}' awarded to the University of Alabama in Huntsville. TT acknowledges the support of the Academy of Finland grant no. 339127.

 This research has made use of the NASA/IPAC Extragalactic Database, which is funded by the National Aeronautics and Space Administration and operated by the California Institute of Technology.

{ MB gratefully acknowledges the years-long support provided by Jukka Nevalainen towards the planning and completion of this project, and regrets
his inability to join this paper and \paperOne\ as a co--author due to other commitments.}

\section*{Data Availability Statement}
All data contained in Tables~\ref{tab:tauEWN-0}--\ref{tab:tauEWN-7}  are provided in full length in the on--line version of the paper, and also in
machine--readable format.  Similar tables for the results of the fits were provided as part of \paperOne.

\bibliographystyle{mnras}

\appendix

\section{{ Curves of growth and tables}}
\label{app:tables}
\label{sec:cog}
{
In 
this
appendix we review the
theory of formation of absorption lines and the relationships between the optical depth at line center $\tau_0$,
the equivalent width of the absorption line $\Wl$, and the column density $N$, following 
the treatment of \cite{draine2011}. 

For a typical Voigt line profile, i.e., the convolution of intrinsic--broadening Lorentz profile and a thermal--broadening Gaussian profile, the optical depth
at line center $\tau_0$ is related to the ion column density $N$ via
\begin{equation}
    \tau_0=\dfrac{\sqrt{\pi} e^2}{m_e c} \dfrac{f \lambda_0}{b} N
    \label{eq:tauoDraine}
\end{equation}
where $b$ is the speed of the ion (both thermal and non--thermal, generally unknown), $f$ is the oscillator strength, $\lambda_0$ the rest--frame
wavelength of the line, and the other symbols have their usual meaning. In our \spex\ fits of \paperOne, the $\tau_0$ parameter is obtained from
a simple Gaussian profile, and therefore it is a phenomenological parameter that is asymptotically equal to that of \eqref{eq:tauoDraine} only for optically--thin absorption lines where the Lorentzian wings are not important.~\footnote{This is also noted
in the \spex\ model page at \texttt{https://spex-xray.github.io/spex-help/models/line.html}. Our data do not have the resolution to fit a more complex line model.} 

The equivalent width $\Wl$ of an absorption
line is empirically defined as the width of a fully absorbed  rectangular line that
has the same integrated transmission as the original line. The dimensionless
equivalent width $W=\Wl/\lambda_0$ is related to the optical depth at line
center in the optically thin--regime ($\tau_0 \ll 1$) by
\begin{equation}
    W = \dfrac{\sqrt{\pi} b}{c} \left(1-\dfrac{\tau_0}{2 \sqrt{2}}\right)\,
    \text{ or } W = \dfrac{\sqrt{\pi} b}{c} \tau_0 \left(1+\dfrac{\tau_0}{2 \sqrt{2}}\right)^{-1}.
    \label{eq:Wthin}
\end{equation}
In this regime, ignoring the correction term in parenthesis in \eqref{eq:Wthin},
there is a linear relationship between the column density and the equivalent width of
the line,
\begin{equation}
    N = \dfrac{m_e c^2}{\pi e^2} \dfrac{1}{f \lambda_0} W\,
    \text{ or } N= 1.13 \times 10^{20} \dfrac{W_{\lambda}}{\lambda_0^2 f} \quad (\text{cm}^{-2}),
    \label{eq:COGLinear}
\end{equation}
as in \cite{draine2011}, see also Eq.~1 of \cite{bonamente2016}. Eq.~\ref{eq:COGLinear} only apply
in the optically--thin regime.

The equivalent width $W_{\lambda}$ of a line is automatically
calculated by \texttt{SPEX} according to its empirical definition and regardless of
assumption on optical depth of the line (see e.g. the \texttt{line} model
component definition and the \texttt{awg} parameter). Our fits therefore also provide the
$\Wl$ parameter that corresponds to the $\tau_0$ parameter, for all absorption--line features. For those fits with emission--line features, i.e.,
when $\tau_0\leq0$, the equivalent width is not defined.

The relationship between the equivalent width and the absorbing
column are usually referred to as the \emph{curve of growth} (COG) of the line, and in general they deviate
from the simple linear relationship given in \eqref{eq:COGLinear} when $\tau_0$ is not
a small number. \cite{rodgers1974} provided a simple formula that approximates the relationship between the $\tau_0$ and $\Wl$ parameters of the form
\begin{equation}
    \dfrac{\Wl}{\lambda_0}= \begin{cases}
     \dfrac{\sqrt{\pi} b}{c} \dfrac{\tau_0}{1+\tau_0/2 \sqrt{2}},\, \text{ for } \tau_0\leq 1.25\\[10pt]
     \dfrac{2 b}{c} \sqrt{\ln{\tau_0/\ln{2}} +\dfrac{\gamma \lambda_0}{4 b \sqrt{\pi}} (\tau_0-1.25)},\, \text{ for } \tau_0>1.25
     \label{eq:COGRodgers}
    \end{cases}
\end{equation}
where $\gamma$ is the damping constant of the Lorentzian profile \citep[e.g.,][]{li2022}. Notice that the top equation in \eqref{eq:COGRodgers} leads to the same result as \eqref{eq:COGLinear} when $\tau_0\ll1$. Equations~\ref{eq:COGRodgers}
provide a simple analytic approximation that interpolates among the  linear, flat (or logarithmic) and damped portions of the curves of growth  \citep{draine2011}.
The system of equations \eqref{eq:tauoDraine} and \eqref{eq:COGRodgers} provides the $N=N(\Wl)$ relationships
that allows the calculation of the column densities implied by the measured equivalent
widths of the absorption lines.

Solution of the system of equations \eqref{eq:tauoDraine}
and \eqref{eq:COGRodgers} can be obtained numerically for all values of $\tau_0$, or equivalently $\Wl$.
Although in general one expects absorption lines from the WHIM to be in the optically--thin regime, the resolution of the data is sometimes so poor that upper limits to the non--detection of absorption lines 
are consistent with large values of $\tau_0$ (and of $\Wl$) that require
a proper curve--of--growth analysis to be converted to upper limits to the underlying
column density. For the \ovii\ and \oviii\ resonance absorption lines under consideration, the damping constants
are respectively $\gamma_{\text{OVII}}=3.43 \times 10^{12}$~s$^{-1}$ and $\gamma_{\text{OVIII}}=2.57\times 10^{12}$~s$^{-1}$ \citep{draine2011, foster2012,foster2020}.~\footnote{These constants are obtained from the Einstein $A$ coefficients of spontaneous emission, and 
can be obtained by the \texttt{ATOMDB} database via the \texttt{pyatombd} software.} 

}

Tables~\ref{tab:tauEWN-0} through~\ref{tab:tauEWN-7} provide a summary of the optical depth $\tau_0$,
the equivalent width $W_{\lambda}$ and the inferred column density $N$ for a fiducial value of $b=100$~\kms.
The second part of the tables are upper limits to these quantities, following the "Feldman-Cousins" or 
the "sensitivity" methods, both giving approximately the same answer. When a `nan' value is present
for the equivalent width or for the column density, that is an indication of a negative value of the optical
depth, which represents an emission-line feature, and therefore these quantities are not defined there.
Only the first ten entries of the tables are displayed. Full tables are reported in the on-line version of the paper. The tables in this paper ignore the few systems where the absorption lines would be in regions that cannot be efficiently measured, such as near the Galactic \ovii\ He-$\alpha$ at 21.6~\AA. Those systems are listed in the on--line ASCII tables.

\begin{table*}
    \centering
    \begin{tabular}{lll|lll|llllll}
    \hline
    \hline
 num & source & z & $\tau_0$ & $W_{\lambda}$ & $\log N$ &  $\tau_{0,FC}^U$ & $\tau_{0,S}^U$ &  $W_{\lambda,FC}^U$ &$W_{\lambda,S}^U$ &  $\log N_{FC}^U$ &$\log N_{S}^U$ \\[5pt]
   &  &  &  & (m\AA) & (cm$^{-2}$) & & &   (m\AA) & (m\AA) &  (cm$^{-2}$)  &  (cm$^{-2}$)  \\
\hline \\
  1 & 1es1028 &  0.12314 & $0.55\pm^{1.74}_{0.88}$ & $4.99\pm^{8.76}_{nan}$  &  $15.305\pm^{0.578}_{nan}$  &2.65 & 2.16 & 16.44 & 15.73 & 16.05 & 16.00 \\
  2 & 1es1028 &  0.13706 & $-0.31\pm^{0.85}_{0.56}$ & $nan\pm^{nan}_{nan}$  &  $nan\pm^{nan}_{nan}$  &0.85 & 1.16 & 7.02 & 9.36 & 15.48 & 15.64 \\
  3 & 1es1028 &  0.33735 & $-0.78\pm^{0.84}_{0.57}$ & $nan\pm^{nan}_{nan}$  &  $nan\pm^{nan}_{nan}$  &0.50 & 1.16 & 4.25 & 9.36 & 15.23 & 15.64 \\
  4 & 1es1553 &  0.18759 & $0.60\pm^{0.32}_{0.26}$ & $5.35\pm^{2.21}_{2.05}$  &  $15.340\pm^{0.182}_{0.238}$  &1.05 & 0.48 & 8.54 & 4.10 & 15.59 & 15.21 \\
  5 & 1es1553 &  0.18775 & $0.57\pm^{0.30}_{0.25}$ & $5.23\pm^{2.16}_{2.08}$  &  $15.329\pm^{0.181}_{0.248}$  &1.00 & 0.45 & 8.13 & 3.90 & 15.56 & 15.18 \\
  6 & 1es1553 &  0.18984 & $-0.05\pm^{0.17}_{0.23}$ & $nan\pm^{nan}_{nan}$  &  $nan\pm^{nan}_{nan}$  &0.26 & 0.33 & 2.34 & 2.90 & 14.94 & 15.04 \\
  7 & 1es1553 &  0.21631 & $0.19\pm^{0.25}_{0.22}$ & $1.94\pm^{2.21}_{nan}$  &  $14.855\pm^{0.359}_{nan}$  &0.57 & 0.39 & 4.82 & 3.37 & 15.29 & 15.11 \\
  8 & 1es1553 &  0.31130 & $-0.15\pm^{0.26}_{0.23}$ & $nan\pm^{nan}_{nan}$  &  $nan\pm^{nan}_{nan}$  &0.27 & 0.40 & 2.42 & 3.50 & 14.96 & 15.13 \\
  9 & 1es1553 &  0.37868 & $0.00\pm^{0.09}_{0.80}$ & $0.03\pm^{0.90}_{nan}$  &  $13.081\pm^{1.441}_{nan}$  &0.72 & 0.73 & 5.98 & 6.10 & 15.40 & 15.41 \\
  10 & 1es1553 &  0.39497 & $0.02\pm^{0.38}_{0.24}$ & $0.18\pm^{3.61}_{nan}$  &  $13.803\pm^{1.366}_{nan}$  &0.50 & 0.51 & 4.27 & 4.36 & 15.23 & 15.24 \\
  \dots\\
  \hline
    \end{tabular}
    \caption{Optical depth, equivalent width and column density for the \ovii/\ovi\ \xmm\ fits 
    (from Table~5 of
    \paperOne), and associated upper limits according to the Feldman-Cousins method ("FC") and the sensitivity 
    method ("S").}
    \label{tab:tauEWN-0}
\end{table*}

\begin{table*}
    \centering
    \begin{tabular}{lll|lll|llllll}
    \hline
    \hline
 num & source & z & $\tau_0$ & $W_{\lambda}$ & $\log N$ &  $\tau_{0,FC}^U$ & $\tau_{0,S}^U$ &  $W_{\lambda,FC}^U$ &$W_{\lambda,S}^U$ &  $\log N_{FC}^U$ &$\log N_{S}^U$ \\
\hline \\
  1 & 1es1028 &  0.13714 & $-0.33\pm^{0.82}_{0.55}$ & $nan\pm^{nan}_{nan}$  &  $nan\pm^{nan}_{nan}$  &0.76 & 1.13 & 6.31 & 9.11 & 15.43 & 15.63 \\
  2 & 1es1028 &  0.20383 & $0.03\pm^{0.97}_{0.83}$ & $0.29\pm^{7.62}_{nan}$  &  $14.014\pm^{1.534}_{nan}$  &1.45 & 1.48 & 11.52 & 11.74 & 15.77 & 15.78 \\
  3 & 1es1028 &  0.22121 & $-0.70\pm^{0.67}_{0.48}$ & $nan\pm^{nan}_{nan}$  &  $nan\pm^{nan}_{nan}$  &0.41 & 0.95 & 3.52 & 7.74 & 15.13 & 15.54 \\
  4 & 1es1553 &  0.03466 & $-0.18\pm^{0.33}_{0.25}$ & $nan\pm^{nan}_{nan}$  &  $nan\pm^{nan}_{nan}$  &0.32 & 0.48 & 2.84 & 4.10 & 15.03 & 15.21 \\
  5 & 1es1553 &  0.04273 & $0.51\pm^{0.64}_{0.44}$ & $4.66\pm^{4.22}_{4.00}$  &  $15.271\pm^{0.343}_{0.902}$  &1.36 & 0.89 & 10.85 & 7.30 & 15.73 & 15.50 \\
  6 & 1es1553 &  0.06364 & $0.53\pm^{0.63}_{0.44}$ & $4.77\pm^{4.11}_{3.88}$  &  $15.282\pm^{0.331}_{0.781}$  &1.41 & 0.88 & 11.15 & 7.24 & 15.75 & 15.50 \\
  7 & 1es1553 &  0.21869 & $-0.01\pm^{0.65}_{0.08}$ & $nan\pm^{nan}_{nan}$  &  $nan\pm^{nan}_{nan}$  &0.59 & 0.60 & 4.98 & 5.07 & 15.30 & 15.31 \\
  8 & 3c249 &  0.13470 & $9.67e+04\pm^{1.00e+20}_{9.68e+04}$ & $46.61\pm^{36.27}_{nan}$  &  $18.239\pm^{0.779}_{nan}$  &8.08e+19 & 8.25e+19 & 1.90e+02 & 1.90e+02 & 19.84 & 19.84 \\
  9 & 3c249 &  0.26664 & $1.70e-01\pm^{1.00e+20}_{3.02e+01}$ & $1.73\pm^{81.15}_{nan}$  &  $14.802\pm^{4.216}_{nan}$  &8.08e+19 & 8.25e+19 & 1.90e+02 & 1.90e+02 & 19.84 & 19.84 \\
  11 & 3c273 &  0.06707 & $0.24\pm^{0.31}_{0.30}$ & $2.34\pm^{2.54}_{nan}$  &  $14.940\pm^{0.354}_{nan}$  &0.74 & 0.50 & 6.14 & 4.29 & 15.41 & 15.23 \\
  \dots\\
  \hline
    \end{tabular}
    \caption{Same as Table~\ref{tab:tauEWN-0}, but for the \ovii/\hi/ \xmm\ fits (from Table~6 of \paperOne).}
    \label{tab:tauEWN-1}
\end{table*}

\begin{table*}
    \centering
    \begin{tabular}{lll|lll|llllll}
    \hline
    \hline
 num & source & z & $\tau_0$ & $W_{\lambda}$ & $\log N$ &  $\tau_{0,FC}^U$ & $\tau_{0,S}^U$ &  $W_{\lambda,FC}^U$ &$W_{\lambda,S}^U$ &  $\log N_{FC}^U$ &$\log N_{S}^U$ \\
\hline \\
  1 & 1es1028 &  0.12314 & $0.02\pm^{3.99}_{0.61}$ & $0.18\pm^{17.69}_{nan}$  &  $14.133\pm^{2.492}_{nan}$  &3.72 & 3.80 & 17.72 & 17.80 & 16.61 & 16.62 \\
  3 & 1es1028 &  0.33735 & $5.01\pm^{15.34}_{3.36}$ & $19.15\pm^{7.90}_{8.13}$  &  $16.749\pm^{0.868}_{0.651}$  &19.83 & 15.43 & 25.61 & 24.23 & 17.46 & 17.30 \\
  4 & 1es1553 &  0.18759 & $0.37\pm^{0.44}_{0.33}$ & $3.57\pm^{3.39}_{3.12}$  &  $15.482\pm^{0.346}_{0.946}$  &0.97 & 0.64 & 7.92 & 5.33 & 15.90 & 15.69 \\
  5 & 1es1553 &  0.18775 & $0.39\pm^{0.49}_{0.31}$ & $3.73\pm^{3.55}_{2.84}$  &  $15.505\pm^{0.349}_{0.662}$  &1.05 & 0.66 & 8.51 & 5.52 & 15.94 & 15.70 \\
  6 & 1es1553 &  0.18984 & $0.98\pm^{0.60}_{0.48}$ & $7.80\pm^{2.97}_{3.21}$  &  $15.893\pm^{0.190}_{0.285}$  &1.85 & 0.89 & 14.44 & 7.30 & 16.33 & 15.86 \\
  7 & 1es1553 &  0.21631 & $-0.08\pm^{0.43}_{0.30}$ & $nan\pm^{nan}_{nan}$  &  $nan\pm^{nan}_{nan}$  &0.52 & 0.60 & 4.39 & 5.07 & 15.59 & 15.66 \\
  8 & 1es1553 &  0.31130 & $1.01\pm^{0.37}_{0.33}$ & $8.36\pm^{2.16}_{2.30}$  &  $15.934\pm^{0.136}_{0.181}$  &1.59 & 0.58 & 12.52 & 4.88 & 16.19 & 15.64 \\
  9 & 1es1553 &  0.37868 & $1.03\pm^{0.44}_{0.35}$ & $8.17\pm^{2.35}_{2.25}$  &  $15.920\pm^{0.149}_{0.180}$  &1.68 & 0.65 & 13.14 & 5.46 & 16.24 & 15.70 \\
  10 & 1es1553 &  0.39497 & $-0.20\pm^{0.19}_{0.18}$ & $nan\pm^{nan}_{nan}$  &  $nan\pm^{nan}_{nan}$  &0.15 & 0.31 & 1.39 & 2.70 & 15.04 & 15.35 \\
  11 & 3c249 &  0.24676 & $4.35e+01\pm^{1.00e+20}_{7.35e+01}$ & $28.32\pm^{55.27}_{nan}$  &  $17.755\pm^{1.888}_{nan}$  &8.08e+19 & 8.25e+19 & 1.90e+02 & 1.90e+02 & 20.44 & 20.44 \\

  \dots\\
  \hline
    \end{tabular}
    \caption{Same as Table~\ref{tab:tauEWN-0}, but for the \oviii/\ovi \xmm\ fits (from Table~7 of \paperOne).}
    \label{tab:tauEWN-2}
\end{table*}

\begin{table*}
    \centering
    \begin{tabular}{lll|lll|llllll}
    \hline
    \hline
 num & source & z & $\tau_0$ & $W_{\lambda}$ & $\log N$ &  $\tau_{0,FC}^U$ & $\tau_{0,S}^U$ &  $W_{\lambda,FC}^U$ &$W_{\lambda,S}^U$ &  $\log N_{FC}^U$ &$\log N_{S}^U$ \\
\hline \\
  2 & 1es1028 &  0.20383 & $0.05\pm^{2.21}_{0.94}$ & $0.50\pm^{13.37}_{nan}$  &  $14.588\pm^{1.701}_{nan}$  &2.55 & 2.60 & 16.30 & 16.38 & 16.48 & 16.49 \\
  3 & 1es1028 &  0.22121 & $-1.47\pm^{0.70}_{0.53}$ & $nan\pm^{nan}_{nan}$  &  $nan\pm^{nan}_{nan}$  &0.19 & 1.01 & 1.70 & 8.24 & 15.13 & 15.93 \\
  4 & 1es1553 &  0.03466 & $-0.03\pm^{0.65}_{0.06}$ & $nan\pm^{nan}_{nan}$  &  $nan\pm^{nan}_{nan}$  &0.54 & 0.59 & 4.57 & 4.94 & 15.61 & 15.65 \\
  5 & 1es1553 &  0.04273 & $0.08\pm^{0.18}_{0.21}$ & $0.88\pm^{1.71}_{nan}$  &  $14.833\pm^{0.495}_{nan}$  &0.39 & 0.32 & 3.42 & 2.83 & 15.46 & 15.37 \\
  6 & 1es1553 &  0.06364 & $0.07\pm^{0.22}_{0.32}$ & $0.70\pm^{2.14}_{nan}$  &  $14.730\pm^{0.641}_{nan}$  &0.49 & 0.45 & 4.19 & 3.83 & 15.56 & 15.52 \\
  7 & 1es1553 &  0.21869 & $-0.43\pm^{0.30}_{0.26}$ & $nan\pm^{nan}_{nan}$  &  $nan\pm^{nan}_{nan}$  &0.14 & 0.46 & 1.32 & 3.96 & 15.02 & 15.53 \\
  9 & 3c249 &  0.26664 & $1.16e+13\pm^{1.00e+20}_{1.16e+13}$ & $74.50\pm^{9.09}_{74.50}$  &  $19.519\pm^{0.125}_{inf}$  &8.08e+19 & 8.25e+19 & 1.90e+02 & 1.90e+02 & 20.44 & 20.44 \\
  10 & 3c273 &  0.00758 & $0.04\pm^{0.42}_{0.05}$ & $0.37\pm^{3.91}_{nan}$  &  $14.456\pm^{1.118}_{nan}$  &0.40 & 0.39 & 3.50 & 3.37 & 15.47 & 15.45 \\
  11 & 3c273 &  0.06707 & $-0.04\pm^{0.44}_{0.08}$ & $nan\pm^{nan}_{nan}$  &  $nan\pm^{nan}_{nan}$  &0.37 & 0.43 & 3.21 & 3.70 & 15.43 & 15.50 \\
  \dots\\
  \hline
    \end{tabular}
    \caption{Same as Table~\ref{tab:tauEWN-0}, but for the \oviii/\hi/ \xmm\ fits (from Table~8 of \paperOne).}
    \label{tab:tauEWN-3}
\end{table*}

\begin{table*}
    \centering
    \begin{tabular}{lll|lll|llllll}
    \hline
    \hline
 num & source & z & $\tau_0$ & $W_{\lambda}$ & $\log N$ &  $\tau_{0,FC}^U$ & $\tau_{0,S}^U$ &  $W_{\lambda,FC}^U$ &$W_{\lambda,S}^U$ &  $\log N_{FC}^U$ &$\log N_{S}^U$ \\
\hline \\
  1 & 1es1028 &  0.12314 & $-0.56\pm^{0.97}_{0.68}$ & $nan\pm^{nan}_{nan}$  &  $nan\pm^{nan}_{nan}$  &0.83 & 1.36 & 6.86 & 10.83 & 15.47 & 15.73 \\
  2 & 1es1028 &  0.13706 & $1.29\pm^{3.09}_{1.34}$ & $10.43\pm^{10.38}_{nan}$  &  $15.709\pm^{0.669}_{nan}$  &4.92 & 3.65 & 18.85 & 17.65 & 16.22 & 16.13 \\
  3 & 1es1028 &  0.33735 & $-0.66\pm^{0.98}_{0.74}$ & $nan\pm^{nan}_{nan}$  &  $nan\pm^{nan}_{nan}$  &0.78 & 1.42 & 6.47 & 11.25 & 15.44 & 15.75 \\
  4 & 1es1553 &  0.18759 & $0.34\pm^{0.57}_{0.46}$ & $3.30\pm^{4.44}_{nan}$  &  $15.102\pm^{0.433}_{nan}$  &1.14 & 0.85 & 9.21 & 6.99 & 15.63 & 15.48 \\
  5 & 1es1553 &  0.18775 & $0.35\pm^{0.62}_{0.47}$ & $3.38\pm^{4.54}_{nan}$  &  $15.113\pm^{0.435}_{nan}$  &1.21 & 0.90 & 9.71 & 7.36 & 15.67 & 15.51 \\
  6 & 1es1553 &  0.18984 & $-0.38\pm^{0.40}_{0.36}$ & $nan\pm^{nan}_{nan}$  &  $nan\pm^{nan}_{nan}$  &0.31 & 0.63 & 2.71 & 5.27 & 15.01 & 15.33 \\
  7 & 1es1553 &  0.21631 & $0.29\pm^{0.58}_{0.46}$ & $2.87\pm^{4.68}_{nan}$  &  $15.037\pm^{0.486}_{nan}$  &1.10 & 0.86 & 8.90 & 7.05 & 15.61 & 15.48 \\
  8 & 1es1553 &  0.31130 & $-0.04\pm^{0.53}_{0.44}$ & $nan\pm^{nan}_{nan}$  &  $nan\pm^{nan}_{nan}$  &0.73 & 0.80 & 6.10 & 6.61 & 15.41 & 15.45 \\
  9 & 1es1553 &  0.37868 & $-0.69\pm^{0.39}_{0.33}$ & $nan\pm^{nan}_{nan}$  &  $nan\pm^{nan}_{nan}$  &0.15 & 0.59 & 1.35 & 5.01 & 14.69 & 15.31 \\
  10 & 1es1553 &  0.39497 & $-0.48\pm^{0.47}_{0.37}$ & $nan\pm^{nan}_{nan}$  &  $nan\pm^{nan}_{nan}$  &0.30 & 0.69 & 2.63 & 5.78 & 14.99 & 15.38 \\

  \dots\\
  \hline
    \end{tabular}
    \caption{Same as Table~\ref{tab:tauEWN-0}, but for the \oviii/\ovi/ \chandra\ fits (from Table~9 of \paperOne).}
    \label{tab:tauEWN-4}
\end{table*}

\begin{table*}
    \centering
    \begin{tabular}{lll|lll|llllll}
    \hline
    \hline
 num & source & z & $\tau_0$ & $W_{\lambda}$ & $\log N$ &  $\tau_{0,FC}^U$ & $\tau_{0,S}^U$ &  $W_{\lambda,FC}^U$ &$W_{\lambda,S}^U$ &  $\log N_{FC}^U$ &$\log N_{S}^U$ \\
\hline \\
 1 & 1es1028 &  0.13714 & $1.34\pm^{2.73}_{1.32}$ & $10.93\pm^{9.89}_{10.73}$  &  $15.737\pm^{0.640}_{1.894}$  &4.50 & 3.34 & 18.48 & 17.31 & 16.19 & 16.11 \\
  2 & 1es1028 &  0.20383 & $1.18\pm^{2.92}_{1.29}$ & $9.52\pm^{9.91}_{nan}$  &  $15.653\pm^{0.613}_{nan}$  &4.47 & 3.47 & 18.45 & 17.46 & 16.19 & 16.12 \\
  3 & 1es1028 &  0.22121 & $0.80\pm^{2.27}_{1.13}$ & $7.05\pm^{9.87}_{nan}$  &  $15.485\pm^{0.597}_{nan}$  &3.43 & 2.80 & 17.41 & 16.66 & 16.12 & 16.06 \\
  4 & 1es1553 &  0.03466 & $-0.03\pm^{0.50}_{0.43}$ & $nan\pm^{nan}_{nan}$  &  $nan\pm^{nan}_{nan}$  &0.70 & 0.77 & 5.87 & 6.35 & 15.39 & 15.43 \\
  5 & 1es1553 &  0.04273 & $0.31\pm^{0.61}_{0.49}$ & $3.08\pm^{4.84}_{nan}$  &  $15.069\pm^{0.479}_{nan}$  &1.17 & 0.91 & 9.38 & 7.43 & 15.65 & 15.51 \\
  6 & 1es1553 &  0.06364 & $-1.05\pm^{0.36}_{0.32}$ & $nan\pm^{nan}_{nan}$  &  $nan\pm^{nan}_{nan}$  &0.10 & 0.56 & 0.98 & 4.75 & 14.55 & 15.28 \\
  7 & 1es1553 &  0.21869 & $-0.47\pm^{0.39}_{0.35}$ & $nan\pm^{nan}_{nan}$  &  $nan\pm^{nan}_{nan}$  &0.22 & 0.61 & 2.02 & 5.14 & 14.87 & 15.32 \\
  8 & 3c273 &  0.00758 & $0.01\pm^{0.82}_{0.62}$ & $0.11\pm^{7.28}_{nan}$  &  $13.582\pm^{1.928}_{nan}$  &1.16 & 1.19 & 9.36 & 9.54 & 15.64 & 15.66 \\
  9 & 3c273 &  0.06707 & $-0.98\pm^{0.55}_{0.47}$ & $nan\pm^{nan}_{nan}$  &  $nan\pm^{nan}_{nan}$  &0.21 & 0.84 & 1.87 & 6.92 & 14.84 & 15.47 \\
  10 & 3c273 &  0.07359 & $2.44\pm^{3.08}_{1.40}$ & $15.79\pm^{6.51}_{6.91}$  &  $16.006\pm^{0.497}_{0.393}$  &6.11 & 3.70 & 19.77 & 17.70 & 16.29 & 16.14 \\

  \dots\\
  \hline
    \end{tabular}
    \caption{Same as Table~\ref{tab:tauEWN-0}, but for the \ovii/\hi/ \chandra\ fits (from Table~10 of \paperOne).}
    \label{tab:tauEWN-5}
\end{table*}

\begin{table*}
    \centering
    \begin{tabular}{lll|lll|llllll}
    \hline
    \hline
 num & source & z & $\tau_0$ & $W_{\lambda}$ & $\log N$ &  $\tau_{0,FC}^U$ & $\tau_{0,S}^U$ &  $W_{\lambda,FC}^U$ &$W_{\lambda,S}^U$ &  $\log N_{FC}^U$ &$\log N_{S}^U$ \\
\hline \\
  1 & 1es1028 &  0.12314 & $-0.79\pm^{0.77}_{0.55}$ & $nan\pm^{nan}_{nan}$  &  $nan\pm^{nan}_{nan}$  &0.47 & 1.09 & 4.00 & 8.80 & 15.54 & 15.96 \\
  3 & 1es1028 &  0.33735 & $-0.24\pm^{1.24}_{0.72}$ & $nan\pm^{nan}_{nan}$  &  $nan\pm^{nan}_{nan}$  &1.29 & 1.62 & 10.28 & 12.71 & 16.06 & 16.21 \\
  4 & 1es1553 &  0.18759 & $0.23\pm^{0.63}_{0.47}$ & $2.25\pm^{4.86}_{nan}$  &  $15.263\pm^{0.578}_{nan}$  &1.11 & 0.91 & 8.97 & 7.43 & 15.98 & 15.87 \\
  5 & 1es1553 &  0.18775 & $0.24\pm^{0.63}_{0.48}$ & $2.36\pm^{4.76}_{nan}$  &  $15.284\pm^{0.557}_{nan}$  &1.12 & 0.92 & 9.04 & 7.49 & 15.98 & 15.87 \\
  6 & 1es1553 &  0.18984 & $0.20\pm^{1.23}_{0.76}$ & $2.01\pm^{8.76}_{nan}$  &  $15.209\pm^{0.874}_{nan}$  &1.81 & 1.64 & 14.11 & 12.89 & 16.31 & 16.22 \\
  7 & 1es1553 &  0.21631 & $-0.45\pm^{0.44}_{0.38}$ & $nan\pm^{nan}_{nan}$  &  $nan\pm^{nan}_{nan}$  &0.29 & 0.68 & 2.57 & 5.65 & 15.33 & 15.72 \\
  8 & 1es1553 &  0.31130 & $-0.05\pm^{0.58}_{0.33}$ & $nan\pm^{nan}_{nan}$  &  $nan\pm^{nan}_{nan}$  &0.69 & 0.75 & 5.75 & 6.23 & 15.73 & 15.77 \\
  9 & 1es1553 &  0.37868 & $-0.37\pm^{0.41}_{0.35}$ & $nan\pm^{nan}_{nan}$  &  $nan\pm^{nan}_{nan}$  &0.31 & 0.63 & 2.71 & 5.27 & 15.35 & 15.68 \\
  10 & 1es1553 &  0.39497 & $-0.63\pm^{0.38}_{0.33}$ & $nan\pm^{nan}_{nan}$  &  $nan\pm^{nan}_{nan}$  &0.18 & 0.59 & 1.64 & 4.94 & 15.12 & 15.65 \\

  \dots\\
  \hline
    \end{tabular}
    \caption{Same as Table~\ref{tab:tauEWN-0}, but for the \oviii/\ovi/ \chandra\ fits (from Table~11 of \paperOne).}
    \label{tab:tauEWN-6}
\end{table*}

\begin{table*}
    \centering
    \begin{tabular}{lll|lll|llllll}
    \hline
    \hline
 num & source & z & $\tau_0$ & $W_{\lambda}$ & $\log N$ &  $\tau_{0,FC}^U$ & $\tau_{0,S}^U$ &  $W_{\lambda,FC}^U$ &$W_{\lambda,S}^U$ &  $\log N_{FC}^U$ &$\log N_{S}^U$ \\
\hline \\
  2 & 1es1028 &  0.20383 & $-1.22\pm^{0.69}_{0.52}$ & $nan\pm^{nan}_{nan}$  &  $nan\pm^{nan}_{nan}$  &0.24 & 1.00 & 2.19 & 8.12 & 15.25 & 15.92 \\
  3 & 1es1028 &  0.22121 & $3.82\pm^{13.01}_{2.75}$ & $17.87\pm^{9.81}_{9.32}$  &  $16.625\pm^{1.061}_{0.678}$  &16.72 & 13.00 & 24.67 & 23.34 & 17.35 & 17.20 \\
  4 & 1es1553 &  0.03466 & $0.21\pm^{0.53}_{0.42}$ & $2.05\pm^{4.29}_{nan}$  &  $15.220\pm^{0.558}_{nan}$  &0.96 & 0.78 & 7.82 & 6.48 & 15.89 & 15.79 \\
  5 & 1es1553 &  0.04273 & $0.00\pm^{0.58}_{0.35}$ & $0.05\pm^{5.22}_{nan}$  &  $13.571\pm^{2.108}_{nan}$  &0.75 & 0.77 & 6.23 & 6.35 & 15.77 & 15.78 \\
  6 & 1es1553 &  0.06364 & $0.86\pm^{0.85}_{0.59}$ & $7.28\pm^{4.53}_{4.64}$  &  $15.854\pm^{0.292}_{0.516}$  &2.04 & 1.19 & 15.52 & 9.54 & 16.42 & 16.01 \\
  7 & 1es1553 &  0.21869 & $-0.35\pm^{0.43}_{0.35}$ & $nan\pm^{nan}_{nan}$  &  $nan\pm^{nan}_{nan}$  &0.32 & 0.64 & 2.78 & 5.39 & 15.36 & 15.69 \\
  8 & 3c273 &  0.00758 & $1.21\pm^{1.27}_{0.81}$ & $9.38\pm^{5.15}_{5.56}$  &  $16.003\pm^{0.335}_{0.487}$  &2.84 & 1.72 & 16.70 & 13.43 & 16.52 & 16.26 \\
  9 & 3c273 &  0.06707 & $0.02\pm^{0.67}_{0.65}$ & $0.18\pm^{5.73}_{nan}$  &  $14.143\pm^{1.597}_{nan}$  &1.07 & 1.09 & 8.63 & 8.80 & 15.95 & 15.96 \\
  10 & 3c273 &  0.07359 & $1.00\pm^{1.61}_{0.87}$ & $7.98\pm^{6.21}_{6.69}$  &  $15.907\pm^{0.406}_{0.898}$  &3.01 & 2.05 & 16.91 & 15.54 & 16.54 & 16.42 \\
  \dots\\
  \hline
    \end{tabular}
    \caption{Same as Table~\ref{tab:tauEWN-0}, but for the \oviii/\hi/ \chandra\ fits (from Table~12 of \paperOne).}
    \label{tab:tauEWN-7}
\end{table*}

\section{Determination of upper limits}
\label{app:UL}

The goal of this appendix is to describe the statistical framework 
to determine upper limits, at a given confidence level, for the $\tau_0$ parameter and therefore for the column density $N$. Upper limits will be indicated with the superscript `U'.
In this paper, we simply use upper limits to set an upper limit
to the cosmological density of baryons according to Eq.~\ref{eq:OmegaWHIMXUL}.
 We are therefore
not interested in a regression of the $N_{\text{ion},i}^U$ measurements to obtain, e.g., their
mean value or other properties. Such methods of regression with upper limits, if interesting
for other applications,
could be performed with the \cite{kaplan1953} estimator, which was
popularized for astronomy by \cite{feigelson1985} as a means to fit data with upper limits.

\subsection{Confidence intervals on model parameters}
\label{sec:confidenceIntervals}
The task of estimating one or more model parameters $\theta$, in this case the single
parameter $\tau_0$ for the line component, is a central problem in statistics.
The point estimate is obtained via the method of \ml, which was pioneered by
\cite{fisher1925}, and it is the method that 
we follow via the minimization of
the \gof\ \cmin\ statistic. The determination of 
confidence intervals on model parameters consists of
devising a suitable method to `invert' probabilities, i.e.,
converting a probability of the observed or estimated
quantity (say, $\hat{\theta}$)
to a probability statement regarding the true model parameter (say, $\theta$). All the considerations provided in the following apply to the general case of a confidence interval 
with both a lower and an upper limit, although they are illustrated here for an upper limit, given the case at hand. Textbook reviews on this topic are provided, for example, in \cite{james2006} or \cite{wasserman2010}.

Let $x$ be the one of the observations of the unknown parameter $\theta=\tau_{0}$, i.e.,
one of the entries in Tables~\ref{tab:tauEWN-0} through \ref{tab:tauEWN-7}.
In general, the observations could be of a different {statistic} $x$ that is related to 
the unknown true parameter $\theta$, although in this case it is a measurement of the parameter itself, via the parametric spectral fits described in the paper. 
The true and unknown parameter $\theta$ is \emph{fixed}, while
the statistic
$x$ is a \emph{random variable} that is assumed to have a known
sampling distribution, and
for which it is therefore meaningful to calculate probabilities of occurrence. In the case at hand, for example, consider
\def\P{\mathrm{P}}
\begin{equation}
\P\left(x \leq x^U | \theta\right) = 1-\alpha
\label{eq:PDirect}
\end{equation}
with $\alpha$ a small number, say $\alpha=0.05$ \citep[e.g.][but other choices are also reasonable]{asa2016}.
The number $x^U$ is an upper limit with confidence $1-\alpha$, say 95\% confidence, meaning 
that,
given the sampling distribution of the statistic, 95\% of the observations should yield a value
that is lower than $x^U$. The notation used for the probability means that one has to assume a
value for the true parameter $\theta$, which is of course unknown.
Often the distribution of a measured
statistic is normal,
and this is the assumption we make in this case; same considerations apply for other distributions.

An upper limit (or more generally, a confidence interval) on a parameter, on  the other hand, can be defined
as the value $\theta^U=\theta^U(x)$, which is a function of the measurement $x$, such that
\begin{equation}
    \P\left(\theta \leq \theta^U(x)\right) = 1-\alpha
    \label{eq:PInverse}
\end{equation}
which is misleadingly similar in form to \eqref{eq:PDirect}, but with a substantially
different meaning. In fact, what is random
here is the { interval}
\begin{equation}
    S(x) = (-\infty, \theta^U(x)),
\end{equation} and not $\theta$, which is fixed. 
Therefore the interpretation of
\eqref{eq:PInverse} is that the range $S(x)$ includes (or "covers") the true value $\theta$
in a  fraction $(1-\alpha)$ 
of cases when such experiment is repeated. 
A clear an concise textbook explanation of this method is also provided in Ch.~11 of~\cite{rohatgi1976},
and a more theoretical discussion in Ch.~12 of \cite{wilks1962}, which summarizes
the original \cite{neyman1937} method for confidence interval estimation. 
In general,
when the parameter $\theta$ is a function of different statistic $x$, 
the data range or set $X^U$ (i.e., $x \in X^U$) corresponds to
a random range $\theta^U(X^U)$ for the parameter,
see e.g. \cite{wilks1962}, and this notation can be extended to more than one
parameter.

\subsection{The Feldman--Cousins method}
\label{sec:FC}
For applications in the physical sciences, often an additional complication
arises that is concerned with certain parameter values being `unphysical', for a given application.
For example, a physical restriction on an absorption line is that the true optical depth parameter
be $\tau_o \geq0$.
Consider as an example one of the measurements from Table~\ref{tab:tauEWN-0},
say $\tau_0=-3.86^{+0.55}_{-0.40}$  in line 17. A negative
best--fit value means that there is evidence for an \emph{emission line}
feature at that wavelength. This feature is likely caused by a random fluctuation of the
Poisson spectrum at that wavelength, or to the presence of an unrelated emission
phenomenon, or perhaps a combination of both. Either way, a formal confidence interval
on the measured parameter $\tau_0$ would remain negative, which would make it impossible to answer the question at hand, which is to set an upper limit to the optical depth on an absorption line,
which needs to be positive.

The seminal paper by \cite{feldman1997} addresses this exact issue, whereas our investigation requires that the true parameter
is non--negative. The
idea is to consider confidence intervals on the unknown parameter of the 
absorption line process, which we continue to indicate as $\theta$ to distinguish it from the measurement itself, based on the ratio of two probabilities,
\begin{equation}
    R = \dfrac{P(x=\tau_0|\theta)}{P(x=\tau_0|\theta_{\mathrm{best}})}
    \label{eq:FC}
\end{equation}
with $\theta_{\mathrm{best}}=\max(0,\tau_0)$. Possible values of the unknown parameter $\theta$ are then chosen among those with the largest
value of the likelihood ratio $R$, subject to the constraint $\theta \geq 0$, while the measurement $\tau_0$
can in fact be negative. This method, hereafter referred to as the FC method, 
requires the specification of the conditional distributions
in \eqref{eq:FC}, which in this case we assume to be Gaussian, and it returns
a range ($\theta_L,\theta^U$) as a confidence interval at a given level of confidence.~\footnote{
For Gaussian data, the method is explained in detail in Sec.~IV~B of \cite{feldman1997}. For \texttt{python}, this method
is implemented in the \texttt{gammapy} software version $\leq 0.7$ (it was somehow removed in later
versions), or from 
the \texttt{FCpy} software \citep{groopman2023}. }

To implement the FC confidence interval method, we start with the assumption that the
probability distribution $P({ x}|\theta)$ is normal. This assumption is somewhat at odds
with the asymmetric errors provided by the $\Delta C=1$ method for several sources, which indicate a more
complex probability model than the Gaussian. In fact, a normal
distribution for the parameter $\tau_0$ would result in a parabolic
log--likelihood function, and therefore symmetric $\Delta C$ intervals 
\citep[see, e.g., Sec.~2.3 of][]{james2006}.
 Nonetheless, we proceed with the Gaussian assumption as an approximation,
and set the standard deviation of the distribution to the average of the
two errors (e.g., we assume a standard deviation $\sigma=0.47$ for the measurement $\tau_0=-3.86^{+0.55}_{-0.40}$), which is thereafter considered as fixed parameter. 
Our crude method to average the positive and negative error
bars is simply one of convenience, given the size of the sample. We expect that the errors
introduced by this method for a given upper limit would average down when inferences are made
for the entire sample.
Moreover, we also adjust the best--fit value to the midpoint of the range (e.g., $\tau_{0,\mathrm{adj}}=-3.79$).
Upon division by the assumed $\sigma$ parameter, all scaled measurements $x=\tau_{0}/\sigma$ 
are therefore  assumed to be distributed like a Gaussian with unit variance, 
$x \sim N(\mu=\theta, \sigma^2=1)$, and to
these scaled values we can then apply the FC method.

\begin{figure}
    \centering
    \includegraphics[width=3.2in]{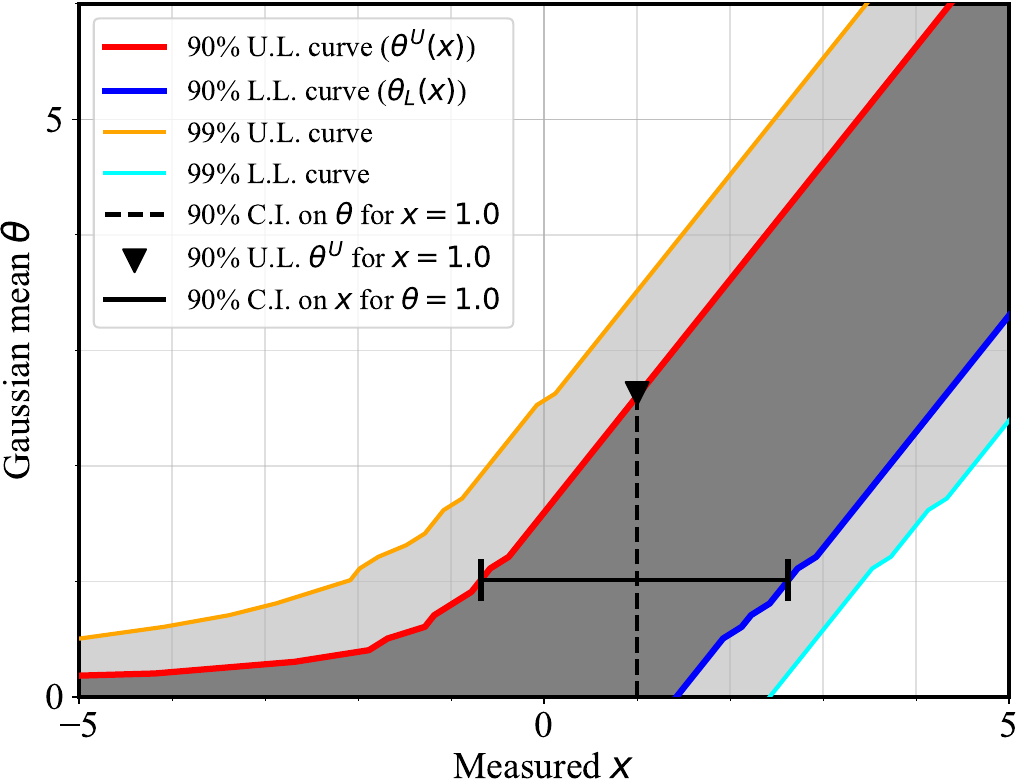}
    \caption{The Feldman--Cousins 90\% and 99\% confidence band for Gaussian measurements $x$, with unit variance ($\sigma^2=1$) and mean $\mu$. A horizontal line between the two curves at a given value of $\mu$ is a 90\% confidence interval on possible measurements drawn from that distribution. A vertical interval between the two curves at the value of a possible measurement $x$ is a 90\% confidence interval on the possible parent  mean $\mu$.}
    \label{fig:FC}
\end{figure}

The situation is illustrated in Fig.~\ref{fig:FC}, which reproduces closely the
example provided by Fig.~10 of  \cite{feldman1997}. The band is constructed horizontally for a given value of the mean $\mu=\theta$,
by requiring
\[
\int_{x_L}^{x^U} P(x|\theta) dx = 1- \alpha{;}
\]
with the choice of $\alpha=0.1$, 
a horizontal range contains 90\% of probability.
In general, there is no unique solution to this problem, i.e., an infinite number of intervals
$(x_L,x^U)$ can be constructed that satisfy this condition.
The FC method requires that the limits $x_L$ and $x^U$ are chosen in such a way that 
the values $ x_L \leq x \leq x^U$ are those that maximize the ratio $R$ according to \eqref{eq:FC}, which contains the assumption that the parent mean must be non--negative.
For a given standardized measurement $x=\tau_{0}/\sigma$, the upper limit
of interest $\theta^U$
is the intersection of the vertical line at $x$ with the red upper limit curve.
Even in those cases with a sufficiently large value of $x$, we only consider an upper limit $\theta^U$, and ignore the lower limit, as a further conservative measure. The FC method 
therefore yields an upper limit to the line parameter $\tau_{0}^U=\theta^U \, \sigma$,
which is then immediately translated into $N^U$.


\subsection{The `sensitivity' method}
\label{sec:ULApproximate}

An alternative and simpler 
method to set upper limits is based
on the following considerations.
First, the best--fit value itself
(i.e., $\tau_0=-3.86$ for the sample entry) is not interesting, 
since a best--fit value that is positive or negative simply means that the center of the
possible line falls at a position with a number of detected counts that is (randomly) either 
lower or greater than the neighboring points.
Second, the most useful part of the regression is the measurement of the parameter's uncertainty, which reflects the ability of a given set of data to constrain the model, or its `sensitivity'.~\footnote{The term 'sensitivity' is used in a similar context also in \cite{feldman1997} to refer to zeroing-out a signal that is not significant, and using estimates of the background to infer the observability of signal.} In fact, a similar fit to a \texttt{line} model in a neighboring portion
of the spectrum would necessarily have a different best--fit parameter $\tau_0$, but a similar uncertainty, since the data have 
similar number of counts, and the null model (i.e., the power--law) is smooth over the 
$\pm 1$~\AA\ range under consideration.

Same as for the FC method, the interest is in setting an upper limit to $\theta$, akin to what is usually done in
source detection with Poisson statistics \citep[e.g.][]{li1983, gehrels1986}. In this case, however, 
we assume that the sampling distribution of the parameter $\tau_{0}$
is a Gaussian 
with unknown mean $\theta \geq 0$, and a variance given by the measured variance $\sigma^2$, same as in the previous method. 
According to this assumption, the one--sided confidence range on $\theta$ with significance
$1-\alpha$ is set by the condition
\begin{equation}
    P(x>0|\theta) = 1-\alpha,
    \label{eq:UL}
\end{equation}
which means that for a mean of $\theta$ there is a small probability $\alpha$ of a negative value of $x$.
It is easy to see that, given the normal sampling distribution, \eqref{eq:UL} is satisfied by
values of $x$ that are, for $\alpha=0.1$ as an example,
\begin{equation}
    x \geq \theta - 1.65\cdot \sigma.
    \label{eq:CIDirect}
\end{equation}


In order to derive a constraint on the parameter $\theta$ itself, it is necessary to look at this confidence interval in a different way. Eq.~\eqref{eq:UL}  implies
\begin{equation}\begin{aligned}
     P\left(\dfrac{x - \theta}{\sigma} \geq - 1.65  \right)= & P(\theta \leq x+1.65\cdot \sigma) = \\ 
     &  P(\theta \leq \theta^U) = 1- \alpha,
     \end{aligned}
    \label{eq:UL2}
\end{equation}
with all $\theta$ values satisfying \eqref{eq:UL2} being the values in this confidence 
interval for the parameter $\theta$, and for a given measurement $x \geq 0$. 
The one--sided confidence interval on $\theta$ is therefore
\begin{equation}
S(x) = (-\infty, 
x+1.65\, \sigma),
\label{eq:CIClassical}
\end{equation}
and it is also referred to as a \emph{fiducial} confidence interval \citep[e.g.][]{fisher1935, wilks1962}.
~\footnote{Further,
\cite{fisher1935} attempted to introduce the concept of \emph{fiducial probability distribution}
on the parameter $\theta$, leveraging the formal similarity between \eqref{eq:UL} and \eqref{eq:UL2}
\citep[see also discussion in Sec. 12.4 of ][]{wilks1962}. Such concept has not gained general acceptance, 
also given certain difficulties with multi--variable problems \citep[see, e.g.][]{tukey1957}.}
The confidence interval $S(x)$ necessarily depends on the measurement $x$ \citep[see, e.g., textbook examples in ][]{rohatgi1976, bonamente2023}. 
If we take the smallest value allowed, $x=0$,
as speculated at the beginning of the section, then 
\eqref{eq:CIClassical} defines the "sensitivity" upper limit.

In general, using $z_{\alpha}$ as the one--sided z--score associated with
a residual probability $P(z \geq z_{\alpha})=\alpha$ (e.g., for $\alpha=0.1,0.01, 0.001$, 
$z_{\alpha}=1.65, 2.58, 3.29)$, we can generalize these argument and
set the $1-\alpha$ sensitivity upper limit to the Gaussian parameter $\theta$ as
$\theta \leq z_{\alpha}\, \sigma$.
This equation defines 
\begin{equation}
    \theta^U = z_{\alpha}\, \sigma
\end{equation} 
as the `sensitivity' upper limit,
at a given confidence level specified by the parameter $\alpha$.

It is useful to notice that the classical confidence interval \eqref{eq:CIClassical}
associated with the actual measurement $x$ could be used instead, given that all such measurements
are available. However, as discussed in Sec.~\ref{sec:FC}, a measurement
$x \leq z_{\alpha}\, \sigma$, such as the sample entry considered before, would yield a formal confidence interval that remains negative, and that is not meaningful.
The `sensitivity' method can therefore be viewed as
a simplified FC method, where the actual measurement is ignored and replaced by the null--hypothesis value
of zero, and without the enforcement of the likelihood--ratio ordering.



\end{document}